\documentclass[ALICE,manyauthors]{cernphprep}

\usepackage[comma,square,numbers,sort&compress]{natbib}
\usepackage{hyperref}
\usepackage{lineno}
\usepackage{upgreek}

\newcommand{\krl}{\ensuremath{\kern-0.18em}}
\newcommand{\krr}{\ensuremath{\kern-0.09em}}
\newcommand{\tms}{\ensuremath{\kern-0.1em\times\kern-0.2em}}
\newcommand{\ptt}{\ensuremath{p_{\mathrm{T}}}\xspace}
\newcommand{\pttpi}{\ensuremath{p_{\mathrm{T}}^{\pi}}\xspace}

\newcommand{\pb}{Pb--Pb\xspace}
\newcommand{\au}{Au--Au\xspace}

\newcommand{\rs}[1][2.76~TeV]{\ensuremath{\sqrt{s}=}~#1\xspace}
\newcommand{\rssv}[1][7~TeV]{\ensuremath{\sqrt{s}=}~#1\xspace}
\newcommand{\rsdv}[1][200~GeV]{\ensuremath{\sqrt{s}=}~#1\xspace}
\newcommand{\rso}{\ensuremath{\sqrt{s}}\xspace}
\newcommand{\rsnn}[1][2.76~TeV]{\ensuremath{\sqrt{s_{\mathrm{NN}}}=}~#1\xspace}
\newcommand{\rsnno}{\ensuremath{\sqrt{s_{\mathrm{NN}}}}\xspace}
\newcommand{\rsnndv}[1][200~GeV]{\ensuremath{\sqrt{s_{\mathrm{NN}}}=}~#1\xspace}
\newcommand{\gvc}{\ensuremath{\mathrm{GeV}\krl/\krr c}\xspace}
\newcommand{\gvcc}{\ensuremath{\mathrm{GeV}\krl/\krr c^{2}}\xspace}
\newcommand{\mvc}{\ensuremath{\mathrm{MeV}\krl/\krr c}\xspace}
\newcommand{\mvcc}{\ensuremath{\mathrm{MeV}\krl/\krr c^{2}}\xspace}
\newcommand{\pion}{\ensuremath{\pi}\xspace}

\newcommand{\pim}{\ensuremath{\pion^{-}}\xspace}
\newcommand{\pip}{\ensuremath{\pion^{+}}\xspace}
\newcommand{\pin}{\ensuremath{\pion^{0}}\xspace}
\newcommand{\kk}{\ensuremath{\mathrm{K}}\xspace}

\newcommand{\km}{\ensuremath{\mathrm{K}^{-}}\xspace}
\newcommand{\kp}{\ensuremath{\mathrm{K}^{+}}\xspace}
\newcommand{\ppi}{\ensuremath{\mathrm{p}\kern-0.05em/\krr\pion}\xspace}

\newcommand{\ks}{\ensuremath{\mathrm{K^{*0}}}\xspace}

\newcommand{\ksm}{\ensuremath{\mathrm{K^{*}\krr(892)^{0}}}\xspace}
\newcommand{\rhof}{\ensuremath{\mathrm{\rho(770)^{0}}}\xspace}
\newcommand{\rhos}{\ensuremath{\mathrm{\rho^{0}}}\xspace}
\newcommand{\rhoa}{\ensuremath{\mathrm{\rho(770)}}\xspace}

\newcommand{\ph}{\ensuremath{\upphi}\xspace}
\newcommand{\phm}{\ensuremath{\ph(1020)}\xspace}
\newcommand{\fnm}{\ensuremath{\mathrm{f_{0}\krr(980)}}\xspace}
\newcommand{\ftm}{\ensuremath{\mathrm{f_{2}\krr(1270)}}\xspace}
\newcommand{\omm}{\ensuremath{\mathrm{\omega\krr(782)}}\xspace}
\newcommand{\ksnm}{\ensuremath{\mathrm{K_{S}^{0}}}\xspace}

\newcommand{\kskm}{\ensuremath{\ks\krl/\km}\xspace}
\newcommand{\kskmf}{\ensuremath{\ksm\krl/\kk}\xspace}

\newcommand{\pks}{\ensuremath{\mathrm{p}\kern-0.1em/\ks}\xspace}
\newcommand{\pksm}{\ensuremath{\mathrm{p}\kern-0.1em/\ksm}\xspace}

\newcommand{\pphi}{\ensuremath{\mathrm{p}\kern-0.1em/\krl\ph}\xspace}
\newcommand{\pphim}{\ensuremath{\mathrm{p}\kern-0.1em/\krl\phm}\xspace}

\newcommand{\omphi}{\ensuremath{\Omega\kern-0.05em/\krr\ph}\xspace}
\newcommand{\omxphi}{\ensuremath{(\Omega^{-}+\overline{\Omega}^{+})\kern-0.05em/\krr\ph}\xspace}
\newcommand{\ommphi}{\ensuremath{\Omega^{-}\kern-0.05em/\krr \ph}\xspace}
\newcommand{\ompphi}{\ensuremath{\overline{\Omega}^{+}\kern-0.05em/\krr \ph}\xspace}
\newcommand{\omphim}{\ensuremath{\Omega\kern-0.05em/\krr\phm}\xspace}

\newcommand{\xipi}{\ensuremath{\Xi\kern-0.1em/\krr\pion}\xspace}
\newcommand{\ompi}{\ensuremath{\Omega\kern-0.05em/\krr\pion}\xspace}
\newcommand{\dd}{\ensuremath{\mathrm{d}}}
\newcommand{\mpt}{\ensuremath{\langle\ptt\rangle}\xspace}

\newcommand{\raa}{\ensuremath{R_{\mathrm{AA}}}\xspace}
\newcommand{\dndy}{\ensuremath{\dd N\krl/\krr\dd y}\xspace}

\newcommand{\npart}{\ensuremath{\langle N_{\mathrm{part}}\rangle}\xspace}
\newcommand{\ncoll}{\ensuremath{\langle N_{\mathrm{coll}}\rangle}\xspace}
\newcommand{\dnc}{\ensuremath{\dd N_{\mathrm{ch}}\kern-0.06em /\kern-0.13em\dd\eta}\xspace}
\newcommand{\dncr}{\ensuremath{\langle \dnc \rangle^{1/3}}\xspace}
\newcommand{\dedx}{\ensuremath{\dd E\krl/\krr\dd x}\xspace}
\newcommand{\stpc}{\ensuremath{\sigma_{\mathrm{TPC}}}\xspace}
\newcommand{\stof}{\ensuremath{\sigma_{\mathrm{TOF}}}\xspace}
\newcommand{\cn}{\ensuremath{\chi^{2}\krl/\krr n_{\mathrm{dof}}}\xspace}
\newcommand{\effr}{\ensuremath{\varepsilon_{\mathrm{rec}}}\xspace}

\begin{document}%

\begin{titlepage}
\PHyear{2018}
\PHnumber{106}      
\PHdate{9 May}
%

\title{Production of the $\rho$(770)$\boldsymbol{^{0}}$ meson \\ 
in pp and \pb collisions at $\boldsymbol{\sqrt{s_{\mathrm{NN}}}}$~=~2.76~TeV}
\ShortTitle{\rhof in pp and \pb collisions}   

\Collaboration{ALICE Collaboration\thanks{See Appendix~\ref{app:collab} for the list of collaboration members}}
\ShortAuthor{ALICE Collaboration} 

\begin{abstract}
The production of the \rhof meson has been measured at mid-rapidity $(|y|<0.5)$ in pp and centrality differential \pb collisions at \rsnn with the ALICE detector at the Large Hadron Collider. The particles have been reconstructed in the $\rhof\rightarrow\pip\pim$ decay channel in the transverse momentum (\ptt) range $0.5-11$~\gvc. A centrality dependent suppression of  the ratio of the integrated yields $2\rhof/(\pip+\pim)$  is observed. The ratio decreases by $\sim40\%$ from pp to central \pb collisions. A study of the \ptt-differential  $2\rhof/(\pip+\pim)$ ratio reveals that the suppression occurs at low transverse momenta, $\ptt<2$~\gvc. At higher momentum, particle ratios measured in heavy-ion and pp collisions are consistent. The observed suppression is very similar to that previously measured for the \kskmf ratio and is consistent with EPOS3 predictions that may imply that rescattering in the hadronic phase is a dominant mechanism for the observed suppression.
\end{abstract}
\end{titlepage}
\setcounter{page}{2}

\section{Introduction\label{sec:intro}}

Due to its very short lifetime ($\tau\sim1.3$~fm/$c$)  the \rhof meson is well suited to study various properties of the interaction dynamics in nucleon-nucleon and heavy-ion collisions~\cite{PDG}. Previous measurements at LEP~\cite{Acton_ZPHYSC_1992_BE, Ackerstaff_EURPHYSJ_1998_BE, Abreu_ZPHYSC_1995_BE, Bedall_PLB_2009_BE} and RHIC~\cite{Adams_PRL_2004_BE} showed that properties of \rhoa mesons reconstructed in the two-pion decay channel are modified in high-energy hadronic interactions and $e^{+}e^{-}$ annihilation. At low momentum, reconstructed \rhoa meson peaks were found to be significantly distorted from the $\textit{p}$-wave Breit-Wigner shape. The observed modifications in the $\rhof \rightarrow \pip\pim$ channel were explained by rescattering of pions ($\pip\pim\rightarrow\rhof\rightarrow\pip\pim$), Bose-Einstein correlations between pions from \rhof decays and pions in the surrounding matter, and interference between differently produced $\pip\pim$ final states~\cite{Abreu_ZPHYSC_1994_BE, Lafferty_ZPHYSC_1993_BE, Bedall_ACTA_2008_BE}. In general, the masses of \rhof mesons produced in hadronic interactions were measured to be systematically lower than the masses measured in $e^{+}e^{-}$ annihilation and a world-averaged difference of $\sim10$~\mvcc was reported in~\cite{PDG}. It is apparent that these effects depend on the charged pion density in the final state and should also play an important role in proton-nucleus and nucleus-nucleus collisions.

In heavy-ion collisions, properties of \rhof mesons can additionally be affected by the hot and dense matter produced in such collisions and by pseudo-elastic or elastic interactions in the late hadron gas stage occuring between chemical and kinetic freeze out. In-medium modification of \rhof mesons was proposed as one of the signals for chiral symmetry restoration~\cite{Petreczky_JPG_2012,Dominguez_PRD_2012,Rapp_ADVPH_2000}. Dilepton continuum measurements in heavy-ion collisions at the SPS~\cite{Agakichiev_PRL_1995,Agakichiev_PLB_1998,Adamova_PLB_2008,Adamova_PRL_2003,Arnaldi_PRL_2006,Arnaldi_PRL_2008,RArnaldi_EPJ_2009,Arnaldi_EPJ_2009} and RHIC~\cite{Adamczyk_PRC_2015,Huang_ACTA_2012,Adare_PRC_2016} indeed exhibit an excess of low-\ptt dilepton pairs below the mass of the \rhof with respect to a hadronic cocktail from all known sources. Results at the SPS and RHIC  are well reproduced by models, which assume that \rhof mesons are regenerated via $\pip\pim$ annihilation throughout the hadron fireball lifetime and freeze out later than the other, longer-lived hadrons. The low-mass dilepton excess is thus identified as the thermal radiation signal from the hadron gas phase, with broadening of the \rhof meson spectral function from the scattering off baryons in the dense hadronic medium and thermal radiation from the QGP. In heavy-ion collisions, rescattering and regeneration are expected to occur between chemical and kinetic freeze-out, affecting the final state yields and peak shapes of short-lived resonances~\cite{Bleicher_JPG_2004,Bleicher_PLB_2002,Knospe_PRC_2016}. Rescattering of daughter particles with the surrounding hadrons changes the kinematics of the decay and some of the resonances can no longer be reconstructed. However the process of regeneration, in which pseudo-elastic scattering of hadrons results in the production of resonances, tends to increase the yields. The cumulative effect depends on the lifetime of the hadronic phase and that of the resonance, as well as on particle cross sections and medium density. Previous measurements at RHIC and the Large Hadron Collider (LHC) showed suppressed production of \ksm~\cite{Abelev_PRC_2015,Adams_PRC_2005} and $\Lambda$(1520)~\cite{Abelev_PRL_2006} but no effect for longer-lived resonances such as the \phm~\cite{Abelev_PRC_2015,Adams_PLB_2005} and $\Sigma$(1385)$^{\pm}$~\cite{Abelev_PRL_2006} in central heavy-ion collisions. These results are qualitatively consistent with expectations from rescattering and regeneration in the hadronic phase. These measurements allowed for model-dependent estimates of the hadronic phase lifetime of at least $2-4$~fm/$c$ in central collisions~\cite{Abelev_PRC_2015,Abelev_PRL_2006}. With the addition of the very short-lived \rhof meson to this study, one can gain additional insight into processes occurring in the late hadronic phase. A measurement of \rhof mesons at high \ptt also contributes to the systematic study of parton energy loss via a measurement of leading hadron suppression~\cite{Bjorken:1982tu,Gyulassy:1990ye,Baier:1994bd}.

The measurement of $\rhof\rightarrow\pip\pim$ in heavy-ion collisions was done only in peripheral \au collisions at \rsnndv, where the ratio of integrated yields, $2\rhof / (\pip+\pim)$  was found to be consistent with that in pp collisions and the reconstructed mass of the \rhof was shifted to lower values~\cite{Adams_PRL_2004_BE}. In this paper, production of \rhof mesons is studied in the $\rhof\rightarrow\pip\pim$ decay channel in pp and centrality differential \pb collisions at \rsnn, including in the 0-20\% most central \pb collisions. Measurements in the hadronic decay channel do not have enough sensitivity for a detailed study of the reconstructed \rhof meson peak shape. As a result, particle yields can be extracted only by using a certain peak model with a limited number of parameters. At present, there are no measurements for the \rhof meson yields or line shapes available in the di-lepton decay channels at LHC energies. Besides, sensitivity to the in-medium spectral function of the \rhof is expected to be different in the di-lepton and the hadronic decay channels. Measurements in the di-lepton channels are sensitive to the whole evolution of the system since leptons leave the fireball mostly unaffected. Measurements in the hadronic channel, because of rescattering and regeneration, should be more sensitive to \rhof mesons, which decay late in the evolution of the hadron gas, where the medium density is low and the mean free path of the decay pions is large. Prediction of the \rhof peak shape in the hadronic channel should rely on the  models that describe the full dynamics of heavy-ion collisions, including the late hadronic phase. An example of such studies performed for \ksm can be found in~\cite{Ilner:2017tab}. Similar studies are not yet available for \rhof. In this work, the yields of \rhof mesons in pp collisions in different \ptt bins were extracted by using a $\textit{p}$-wave relativistic Breit-Wigner function corrected for phase space, a mass dependent reconstruction efficiency and pion interference as described by the S{\"o}ding parameterization~\cite{Soding_PL_1966}. The peak position was kept as a free parameter. Due to the lack of detailed predictions for the \rhof meson peak shape as a function of transverse momentum and centrality in heavy-ion collisions, the same model was also used in \pb collisions.

The paper is organized as follows. Details of the data analysis and the peak model are described in Section 2. Sections 3 and 4 present details on the normalization and corrections used to obtain the invariant differential yields of \rhof mesons in pp and \pb collisions. Results, including \rhof meson yields, reconstructed masses, particle ratios and nuclear modification factors are presented in Section 5 and compared to model predictions where available. For the remainder of this paper, the \rhof will be denoted by the symbol \rhos and the half sum of the charged pion yields $(\pip+\pim)/2$ as $\pi$.

\section{Data analysis\label{sec:analysis}}

\subsection{Event and track selection\label{sec:analysis:event}}

In this work, the production of \rhos mesons is measured at mid-rapidity $(|y|<0.5)$ in \pb and pp collisions at \rsnn using the data samples collected by the ALICE experiment at the LHC during the 2010 and 2011 data taking periods, respectively. The experimental setup and the event selection criteria for these periods are described in detail in previous ALICE publications on resonance production~\cite{Abelev_PRC_2015,Kishora}. 

The main detector subsystems used in this analysis are the V0 detectors, the Inner Tracking System (ITS), the Time Projection Chamber (TPC) and the Time-of-Flight (TOF) detector~\cite{Aamodt_JI_2008}. The minimum bias trigger in pp collisions was configured to obtain high efficiency for hadronic interactions and required at least one hit in either of the V0 detectors (V0A and V0C) or in the Silicon Pixel Detector (SPD), which constitutes the two innermost layers of the ITS. In \pb collisions, the minimum bias trigger required at least two out of the following three conditions: (i) two hits in the outer layer of the SPD, (ii) a signal in V0A, (iii) a signal in V0C~\cite{Abbas_JINST_2013}. The collision centrality is determined on the basis of the multiplicity measured in the V0 detectors. Glauber-model simulations are used to estimate the average number of participants (\npart) and number of binary inelastic nucleon-nucleon collisions (\ncoll) for each selected centrality interval~\cite{Aamodt_PRL_2011,Abelev_PRC_2013}. The number of analyzed minimum bias events is equal to about $6\times10^{7}$ in pp collisions, corresponding to an integrated luminosity of $\mathcal{L}_{int} = N_{\mathrm{MB}} / \sigma_{\mathrm{MB}} = 1.1$~nb$^{-1}$, where  $N_{\mathrm{MB}}$ and $\sigma_{\mathrm{MB}} = (55.4 \pm 1.0)$~mb are the number and cross section of pp collisions passing the minimum bias trigger conditions~\cite{Abelev_EPJ_2013}. In \pb collisions the number of analyzed events is $17.5\times10^{6}$. The TPC is used to reconstruct charged particle tracks with the requirement that the track has crossed at least 70 read-out rows out of a maximum 159~\cite{Abelev:2014ffa}. Only high-quality tracks reconstructed with the TPC and ITS are selected for analysis; tracks are required to be matched to the primary vertex within 2~cm in the longitudinal direction and within 7$\sigma$ in the transverse plane, where $\sigma$ is (0.0015 + 0.0050/$\ptt^{1.1}$) cm for pp and (0.0026 + 0.0050/$\ptt^{1.01}$) cm for \pb~\cite{Abelev_PRC_2015}, with \ptt in units of \gvc. The primary vertex is required to be within $\pm10$~cm of the detector center along the beam axis. Tracks are required to have a minimum transverse momentum of 150~\mvc in pp collisions and 400~\mvc in \pb collisions and a pseudorapidity of $|\eta|<0.8$. The higher \ptt cut in \pb collisions was needed to improve the signal-to-background ratio at low and intermediate momentum. To be identified as charged pions, reconstructed tracks in pp collisions need to have a specific ionisation energy loss \dedx measured in the TPC within $2\stpc$ of the expected value. For \pb collisions, particles with a signal in the TOF subsystem are identified by requiring the time-of-flight and \dedx to be within $2\stof$ and $5\stpc$ of the expected values, respectively. Particles without a signal in the TOF are identified in the same way as in pp collisions. The $\stpc$ is about 5\% for isolated tracks and 6.5\% for central \pb collisions. The typical value of $\stof$ is about 80 ps.

\subsection{Yield extraction\label{sec:analysis:yield}}

\begin{figure}
\includegraphics[width=19pc]{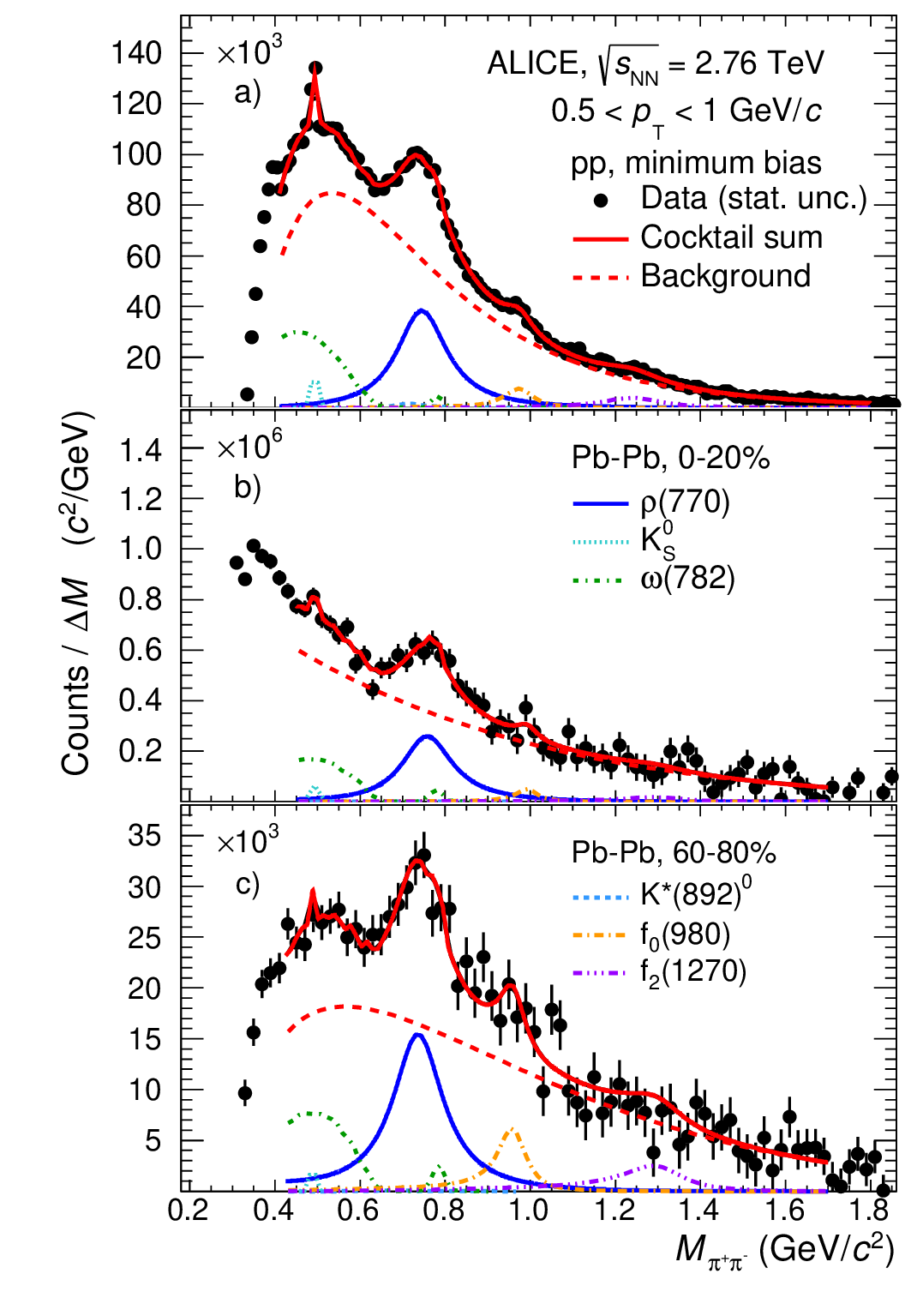}
\includegraphics[width=19pc]{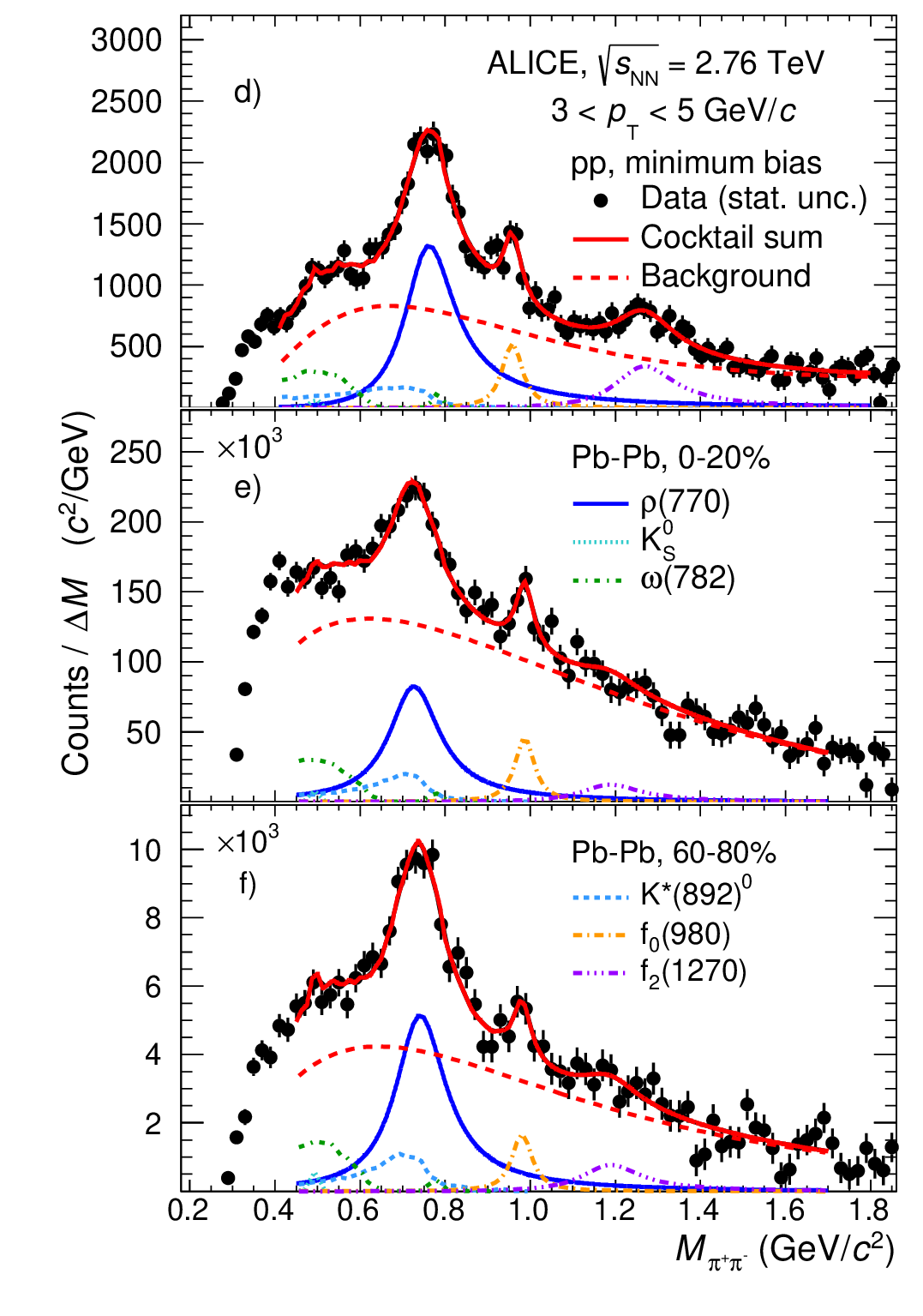}
\caption{\label{fig:analysis:minv}(Color online) Invariant mass distributions for $\pip\pim$ pairs after subtraction of the like-sign background. Plots on the left and right are for the low and high transverse momentum intervals, respectively. Examples are shown for minimum bias pp, 0--20\% and 60--80\% central \pb collisions at \rsnn. Solid red curves represent fits to the function described in the text. Colored dashed curves represent different components of the fit function, which includes a smooth remaining background as well as contributions from \ksnm, \rhos, \omm, \ksm, \fnm and \ftm. See text for details.}
\end{figure}

Yields of \rhos mesons for each \ptt and centrality interval are measured by calculating invariant mass distributions of oppositely charged identified pions ($\pip\pim$ pairs). The combinatorial background is estimated using the like-sign method: this background is $2\sqrt{N^{++}N^{--}}$, where $N^{++}$ and $N^{--}$ are the numbers of $\pip\pip$ and $\pim\pim$ pairs within the same event, respectively. In addition to the uncorrelated combinatorial background, the like-sign method also partly subtracts the minijet~\cite{Sjostrand:1987su} contribution in the background; this is the main reason why it is preferred to the mixed-event approach in this analysis. However, production of like-sign and opposite-sign pairs in jets differ and a perfect background description is not expected. Examples of invariant mass distributions after subtraction of the like-sign background in minimum bias pp, 0--20\% and 60--80\% central \pb collisions at \rsnn are shown in Fig.~\ref{fig:analysis:minv}. The analysis has also been performed using an event-mixing technique to compute the combinatorial background. The \rhos yields obtained using event mixing are consistent with those obtained when a like-sign background is subtracted.

After subtraction of the like-sign background, the resulting distributions contain the remaining correlated background from minijets and pairs from hadronic decays. The latter has a very complex shape, which depends on $\pip\pim$ pair invariant mass and transverse momentum. The main contributions to the correlated background are: (i) $\omm\rightarrow\pin\pip\pim$, $\omm\rightarrow\pip\pim$, $\fnm\rightarrow\pip\pim$, $\ftm\rightarrow\pip\pim$ and $\ksnm\rightarrow\pip\pim$, (ii) $\ksm\rightarrow$$K^{\pm}\pi^{\mp}$, where the charged kaon in the final state is reconstructed as a pion, and (iii) $\eta\rightarrow\pin\pip\pim$, $\eta^{'}$(958)$\rightarrow\eta\pip\pim$ and $\phm\rightarrow\km\kp$ decays. The first two contributions overlap with the wide \rhos meson peak and need to be correctly accounted for as described in Section 2.1.2. The last contribution can be neglected if the analysis is limited to a mass range of $M_{\mathrm{\pip\pim}}> 0.4$~\gvcc. Contributions from misreconstructed decays of heavier hadrons do not result in peaked structures and were estimated to be negligible.

In order to extract the \rhos yields, the invariant mass distributions after subtraction of the combinatorial like-sign background are fitted with a function that accounts for all known correlated contributions to the $\pip\pim$ mass distribution. In this section, we discuss the assumptions used to approximate different components of the background.

\subsubsection{Background from minijets\label{sec:analysis:yield:jets}}

The invariant mass distribution of $\pip\pim$ pairs has been extensively studied using full event Monte-Carlo simulations of the experimental setup. PYTHIA 6~\cite{TSjostrand_JHEP_2006} and HIJING~\cite{Wang_PRD_1991} are used as event generators for pp and \pb collisions, respectively. The produced particles and their decay products are propagated through the ALICE detector using GEANT 3~\cite{Brun_CERN_1994}. Invariant mass distributions for pairs of charged pions are accumulated after application of the same event, track and particle identification cuts as in data. The study shows that after subtraction of the like-sign background and known contributions from \ksnm, $\eta$, \rhos, \omm, \ksm, $\eta^{'}$(958), \fnm, \phm and \ftm, the remaining background has a smooth dependence on mass. Based on a dedicated study of PYTHIA simulations, this remaining contribution is due to minijets. As described in ~\cite{Ackerstaff_EURPHYSJ_1998_BE,Abreu_ZPHYSC_1995_BE,Bedall_PLB_2009_BE}, the remaining background is parameterized with the following function: $F_{\mathrm{BG}}(M_{\mathrm{\pi\pi}})=(M_{\mathrm{\pi\pi}}-2m_{\mathrm{\pi}})^{n} \cdot\mathrm{exp}(A+B\cdot M_{\mathrm{\pi\pi}}+C\cdot M_{\mathrm{\pi\pi}}^{2})$, where $m_{\mathrm{\pi}}$  is the mass of the charged pion and $n$, $A$, $B$ and $C$ are fit parameters. It has been checked that this function describes the remaining background in Monte-Carlo events for all analyzed \ptt and centrality intervals. A lower order polynomial in the exponential would not provide enough flexibility for the function to describe the remaining background in a wide mass range. A higher order polynomial, while not improving the fit quality, could result in unjustified fluctuations of the background function. When a polynomial is tried as a fit function, it needs larger number of fit parameters to describe the background in the same mass range. Parameters of the background function are not constrained in fits to data.

\subsubsection{Contributions from \ksnm, \omm and \ksm \label{sec:analysis:yield:cocktail1}}

The production of \ksm mesons in pp and \pb collisions at \rsnn was measured in~\cite{Abelev_PRC_2015,Kishora}. The yield of \ksnm mesons in pp collisions is estimated as $(\kp + \km)/2$ using the charged kaon measurements published in~\cite{Abelev_PLB_2014}. For \pb collisions, the production of \ksnm mesons was measured in~\cite{Abelev_PRL_2013}. The production of \omm mesons has not been measured in the collision systems under study. However, it has been estimated using procedures similar to those previously used in calculations of hadronic cocktails in the dilepton continuum or direct photon measurements~\cite{Adamczyk_PRC_2015,Adare_PRC_2016,Adler_PRL_2005,Adam_PLB_2016}.

Contributions from \ksnm, \omm and \ksm are approximated with templates extracted from Monte-Carlo simulations and normalized to known yields. The template shapes are simulated by applying the same analysis cuts as in data and reconstructing the \ksnm, \omm and \ksm mass shapes in the $\pip\pim$ channel separately for each \ptt and centrality interval used in the \rhos analysis. Then the templates are normalized to the independently measured \ksnm, \omm and \ksm yields in the corresponding intervals, corrected for the branching ratios and the acceptance times reconstruction efficiency values ($A\times\effr$, hereafter ``efficiency") estimated in simulations.

\begin{figure}
\includegraphics[width=19pc]{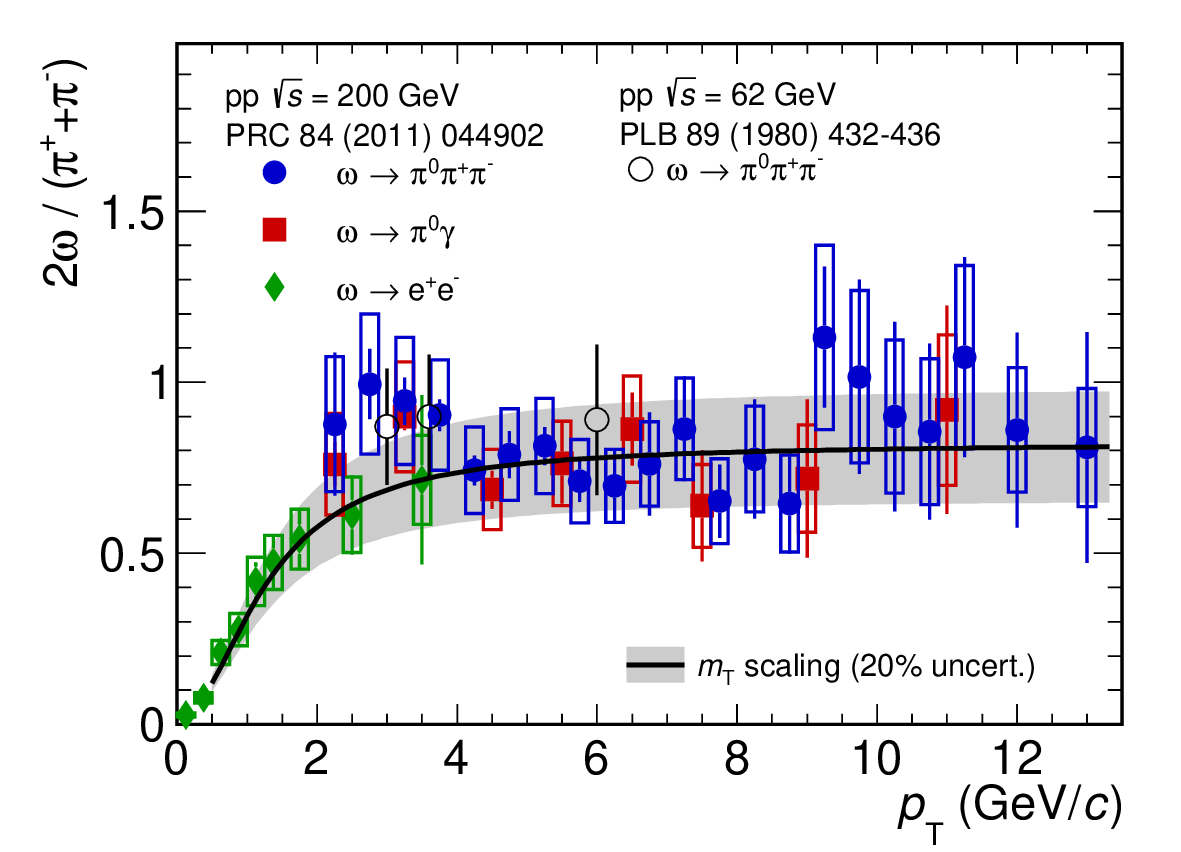}
\includegraphics[width=19pc]{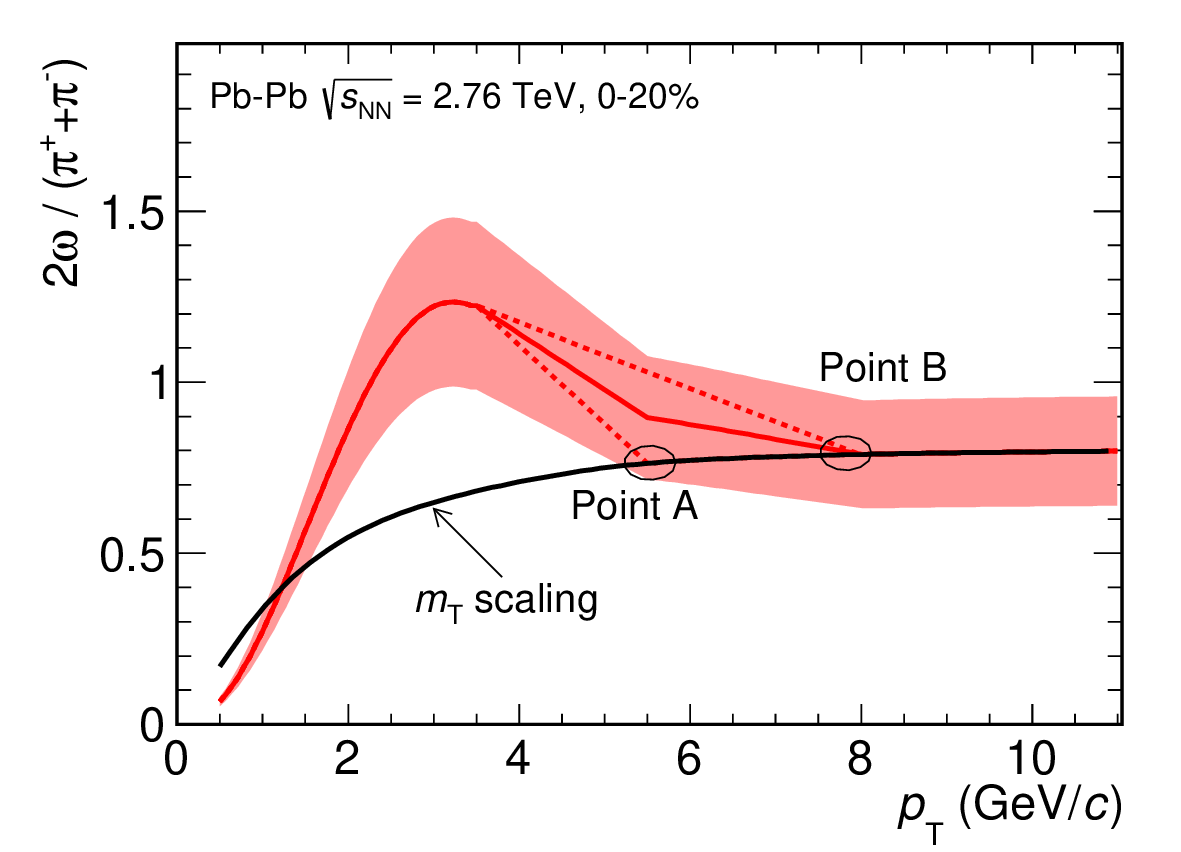}
\caption{\label{fig:analysis:ompi}(Color online) Left: measured $\omega/\pi$ ratio as a function of transverse momentum in pp collisions at $\rso = 62$ and $200$~GeV~\cite{Adare_PRC_2011,Diakonou_CERN_1979}. The smooth curve shows the estimated $\omega/\pi$ ratio in pp collisions at \rs. Right: estimated $\omega/\pi$ ratio as a function of the transverse momentum for 0--20\% central \pb collisions at \rsnn. The shaded regions in the two panels correspond to a $\pm20\%$ variation of the ratio, see text for details.}
\end{figure}

Measurements of the \omm meson \ptt spectrum are rare. A summary of the world-wide measurements of \omm mesons in pp collisions at different energies is given in the left panel of Fig.~\ref{fig:analysis:ompi}~\cite{Adare_PRC_2011,Diakonou_CERN_1979}. The data are presented in terms of the $\omega/\pi$ ratio. Most of the data come from PHENIX measurements at \rsdv. It is important to note that the $\omega/\pi$ ratio does not depend on the collision energy within uncertainties in the range $\rso=62-200$~GeV. In this analysis, it is assumed that the $\omega/\pi$ ratio stays constant in the range $\rso=200-2760$~GeV. This assumption is supported by other light-flavor meson ratios like $K/\pi$, $\eta/\pi$ and $\phi/\pi$, which do not show any significant energy dependence in pp collisions in the range $\rso=200-7000$~GeV~\cite{Abelev_PLB_2014,Adare_PRC_2013,Abelev_EPJ_2012,Aamodt_EPJ_2011,Abelev_PLB_2012,Adler_PRC_2007}. This assumption is also confirmed with PYTHIA 6~\cite{TSjostrand_JHEP_2006} and PYTHIA 8~\cite{Sjostrand_CPC_2008} calculations, which predict the $\omega/\pi$ ratios in pp collisions at $\rso=200$~GeV and $\rso=2.76$~TeV to be consistent within 10\%. 
 
For pp collisions at \rs, the shape of the \omm \ptt spectrum is approximated using the fit to $\pi$ spectrum~\cite{Abelev_PLB_2014} with a $m_{\mathrm{T}}$-scaling correction~\cite{Altenkamper:2017qot}. The function, which is fit to the transverse momentum distribution of pions, is transformed into a production spectrum of \omm in two steps. First, \pttpi of charged pions in the function is replaced with $\sqrt{m_{\mathrm{\omega}}^{2}-m_{\mathrm{\pi}}^{2}+{p_{\mathrm{T}}^{\omega}}^{2}}$, where $m_{\mathrm{\omega}}$ and  $p_{\mathrm{T}}^{\omega}$ are the mass and transverse momentum of the $\omega$ meson. Second, the resulting function is scaled to the $\omega/\pi$ ratio measured at high transverse momentum. Based on the left panel of Fig.~\ref{fig:analysis:ompi}, the ratio is normalized to $\omega/\pi=0.81$, which is in agreement with the value of $\omega/\pi=0.81\pm0.02\pm0.09$ measured by PHENIX in pp collisions at $\rso=200$~GeV~\cite{Adare_PRC_2011}. The ratio of the derived \omm spectrum to the measured $\pi$ spectrum is shown with a curve in the same figure. The shaded region corresponds to a $\pm20\%$ variation of the $\omega/\pi$ ratio, which is used in Section 5 to estimate the systematic uncertainty for measurement of the \rhos yields.
 
For \pb collisions one has to additionally account for radial flow that modifies the shapes of particle production spectra at low and intermediate transverse momenta. The strength of the radial flow for each centrality interval is estimated by simultaneous fits of the charged pion, kaon and (anti)proton production spectra~\cite{BAbelev_PRC_2013,Adam_PRC_2016} with Tsallis Blast-Wave (TBW) functions~\cite{Tang_PRC_2009}. The non-extensivity parameter $q$ is set to be equal for mesons and baryons, thus keeping all fit parameters the same for all particles, except for particle masses and normalizations. Pions, kaons and protons are fit in similar \ptt ranges, from the lowest measured momentum ($0.1$~\gvc, $0.2$~\gvc and $0.3$~\gvc for pions, kaons and protons, respectively) up to $3.5$~\gvc. In this range, the fits reproduce the measured results within the experimental uncertainties. Fits performed in different \ptt ranges, $0.1-2$~\gvc and $0.5-3.5$~\gvc produce very similar results and therefore lead to negligible systematic ucncertainties in this procedure. For all fits, it has been checked that the total integrated yields extracted from the fit curves are consistent with the published values within uncertainties. The expected \omm \ptt spectrum is parameterized with a TBW function with the \omm mass~\cite{PDG} and all other fit parameters set to the values from the combined fit. This function is normalized so that the ratio of the integrated yields $\omega/\pi=0.1$. This value of the ratio was previously measured with high precision in pp and $e^{+}e^{-}$ collisions~\cite{Ackerstaff_EURPHYSJ_1998_BE,Adare_PRD_2011}. Measurements of the ratio in heavy-ion collisions are available only from STAR in \au collisions at \rsnndv: $\omega/\pi=0.086\pm0.019$~\cite{Adamczyk_PRC_2015}. This measurement is in good agreement with pp results. This is similar to the $K$/$\pi$ and $p$/$\pi$ ratios, which vary only within $\sim20\%$ from pp to central heavy-ion collisions for $\rsnno=200-2760$~GeV~\cite{BAbelev_PRC_2013}. The resulting $\omega/\pi$ ratio is shown in the right panel of Fig.~\ref{fig:analysis:ompi} with a solid red curve. The shaded region around the curve corresponds to a $\pm20\%$ variation of the ratio, which is used to estimate the systematic uncertainty (see Section 5). Up to $\ptt=3.5$~\gvc the ratio is determined from TBW fits as described above. It is important to note that two alternative approaches are also used for estimation of the \omm production spectrum in this \ptt range. In these approaches, only the production spectra for charged pions and kaons or only the spectra for charged kaons and (anti)protons are used to fix parameters of the TBW function. Both approaches result in $\omega/\pi$ ratios which are consistent with the default value within the shaded region.

At very high transverse momentum it is assumed that the $\omega/\pi$ ratio returns to the same values measured in pp collisions (a $m_{\mathrm{T}}$-scaled $\pi^{\pm}$ spectrum), similar to what is observed for other ratios like $K$/$\pi$ and $p$/$\pi$~\cite{Abelev_PLB_2014}. This assumption is also confirmed by PHENIX in \au collisions at \rsnndv where the $\omega/\pi$ ratio was measured at high \ptt to be $\omega/\pi=0.82\pm0.09\pm0.06$~\cite{Adare_PRC_2011}, very close to the value in pp collisions. The exact \ptt value at which the influence of radial flow becomes negligible for \omm mesons is not known. It is expected to be mass dependent and sit in between \ptt values where the $K$/$\pi$ (Point A) and $p$/$\pi$ (point B) ratios measured in \pb collisions merge with those measured in pp collisions. The dashed lines show how the $\omega/\pi$ ratio would look in the transition region if the merging point to the $m_{\mathrm{T}}$-scaled curve was the same as for $K$/$\pi$ or $p$/$\pi$. For the nominal $\omega/\pi$ ratio we choose the average of the $\omega/\pi$ ratios obtained for these two extreme cases, shown with a solid line. The merging point for \omm is varied between the merging points for $K$/$\pi$ and $p$/$\pi$ for a study of systematic uncertainties. One can see that the two extreme cases for the transition are within the shaded region. 	

\subsubsection{Contributions from \rhos, \fnm and \ftm \label{sec:analysis:yield:cocktail2}}

Contributions from \rhos, \fnm and \ftm mesons are described analytically. The shapes of these resonances are described with a relativistic Breit-Wigner function (rBW)~\cite{Adler_PRL_2002,Adam_JHEP_2015}:

\begin{equation}
\label{eq:analysis:rbw}
\mathrm{rBW}(M_{\mathrm{\pi\pi}}) = \frac{A M_{\mathrm{\pi\pi}} M_{\mathrm{0}} \Gamma(M_{\mathrm{\pi\pi}}) }{(M_{\mathrm{0}}^{2} - M_{\mathrm{\pi\pi}}^2)^2 + M_{\mathrm{0}}^{2} \Gamma^{2}(M_{\mathrm{\pi\pi}})},
\end{equation}
  
where $M_{\mathrm{0}}$ is the mass of the resonance under study and $A$ is a normalization constant. For wide resonances one should account for the dependence of the resonance width on mass:

\begin{equation}
\label{eq:analysis:gamma}
\Gamma(M_{\mathrm{\pi\pi}}) = \left ( \frac{M_{\mathrm{\pi\pi}}^{2}-4m_{\mathrm{\pi}}^{2}}{M_{\mathrm{0}}^{2}-4m_{\mathrm{\pi}}^{2}}\right ) ^{\mathrm{(2J+1)/2}} \times \Gamma_{\mathrm{0}} \times M_{\mathrm{0}}/M_{\mathrm{\pi\pi}},
\end{equation}
 
where $\Gamma_{\mathrm{0}}$ is the width of the resonance, $m_{\mathrm{\pi}}$ is the charged pion mass and $J$ is equal to 0 for \fnm, 1 for \rhos and 2 for \ftm. The masses of \rhos, \fnm and \ftm are kept as free parameters. As has been pointed out, measurements of \rhos mesons in the hadronic decay channel do not have enough sensitivity for a detailed study of the resonance peak shape. As a result, the width of the \rhos is fixed to $149.3$~\mvcc, which corresponds to the resonance width in vacuum $147.8\pm0.9$~\mvcc~\cite{PDG} convoluted with the detector mass resolution extracted from simulations. Due to the large width of the \rhos peak, its smearing due to the mass resolution results in a negligible change in the extracted yields. The width of the \fnm is limited to be within 40-100~\mvcc and the width of the \ftm is fixed to $186.7$~\mvcc~\cite{PDG}.
 
Since resonances can be produced through $\pi\pi$ scattering in the hadronic phase, the reconstructed peaks can be affected by the phase space available for pions. It was suggested in~\cite{Adams_PRL_2004_BE,Rapp_NPA_2003,Kolb_PRC_2003,Shuryak_NPA717_2003} to use a Boltzmann factor to account for the phase space correction

\begin{equation}
\label{eq:analysis:ps}
PS(M_{\mathrm{\pi\pi}}) = \frac{M_{\mathrm{\pi\pi}}}{\sqrt{M_{\mathrm{\pi\pi}}^{2} + \ptt^{2}}} \times \mathrm{exp}\negthickspace \left ( -\sqrt{M_{\mathrm{\pi\pi}}^{2} + \ptt^{2}}/T \right ),
\end{equation}
 
where $T$ is the kinetic freeze-out temperature, set to $160$~MeV in pp and $120$~MeV in heavy-ion collisions~\cite{BAbelev_PRC_2013,BAbelev_PLB_2014}.

The \rhos, \fnm and \ftm resonances have quite large widths. Efficiencies for these mesons can change with particle masses at a given transverse momentum, resulting in distortion of the reconstructed peak shapes. The effect is most prominent at low \ptt, where the efficiency $A\times\effr$ rapidly increases with mass and transverse momentum. Therefore, the peak shapes for \rhos, \fnm and \ftm are corrected for the dependence of $A\times\effr$ on the particle masses. The corresponding corrections are evaluated from Monte-Carlo simulations.

Previous measurements showed that \rhos meson peaks reconstructed in the $\pip\pim$ decay channel are distorted: the central value (mass) was shifted to lower values by tens of \mvcc. This phenomenon was studied in detail at LEP~\cite{Acton_ZPHYSC_1992_BE,Ackerstaff_EURPHYSJ_1998_BE,Abreu_ZPHYSC_1995_BE,Bedall_PLB_2009_BE} and was also observed at RHIC~\cite{Adams_PRL_2004_BE}. The modification of the reconstructed \rhos meson shape was explained by Bose-Einstein correlations between identical pions in the final state (including decay pions from short-lived \rhos mesons) and interference between final states which are either two directly produced pions or two pions from \rhos decays. Both effects result in a similar modification of the peak shape, which at LEP was accounted for by including an interference term parameterized by S{\"o}ding~\cite{Soding_PL_1966} in the peak model

\begin{equation}
\label{eq:analysis:soding}
f_{\mathrm{i}}(M_{\mathrm{\pi\pi}}) = C \left ( \frac{M_{\mathrm{0}}^{2}-M_{\mathrm{\pi\pi}}^{2}}{M_{\mathrm{\pi\pi}} \Gamma(M_{\mathrm{\pi\pi}})} \right ) f_{\mathrm{s}}(M_{\mathrm{\pi\pi}}),
\end{equation}
 
where $f_{\mathrm{s}}(M_{\mathrm{\pi\pi}})$ is the default peak shape as described above, $f_{\mathrm{i}}(M_{\mathrm{\pi\pi}})$ is the interference term and $C$ is a free parameter that determines the strength of the interference. Using this term in the peak model enhances the left side of the reconstructed peak and suppresses the right side of the peak. If one fits the distorted peak with the regular rBW function, the reconstructed mass is shifted towards lower values and the fit quality is poor due to the distorted tails. We note that RHIC~\cite{Adler_PRL_2002} and LHC~\cite{Adam_JHEP_2015} measurements of photoproduction of \rhos mesons in ultra-peripheral heavy-ion collisions were performed with this S{\"o}ding correction included in the peak model and the reconstructed parameters of \rhos were found to be in agreement with vacuum values. In this study, the extraction of \rhos meson yields is performed using peak models with and without the S{\"o}ding interference term. For the hadronic interactions  the S{\"o}ding correction is just empirical. The peak model with the term somewhat better describes the measured peaks at low momentum and leads to reconstructed meson masses closer to the accepted vacuum value~\cite{PDG} and is used by default. The peak model without the interference term is used in the evaluation of the systematic uncertainties.

In heavy-ion collisions, the shape of the \rhos meson peak can also be distorted due to chiral symmetry restoration~\cite{Petreczky_JPG_2012,Dominguez_PRD_2012,Rapp_ADVPH_2000} in the earlier stages of the collisions and due to rescattering, regeneration, correlations, and interference in the later stages~\cite{Bleicher_JPG_2004,Bleicher_PLB_2002,Knospe_PRC_2016}. The relative strengths of these effects are not well understood and there are no detailed predictions for the \ptt and centrality dependence of \rhos peak modifications that take all of them into account. In this analysis, we therefore limit our peak model to the effects discussed in the preceding paragraphs.

In summary, the default \rhos peak model used in this analysis is the product of a relativistic Breit-Wigner function (with a mass-dependent width), a phase-space factor, a mass-dependent efficiency correction, and a S{\"o}ding interference term. The same peak model, only without the interference term, is used to fit the \fnm and \ftm peaks.

\subsubsection{Fit results\label{sec:analysis:yield:fits}}

The fitting function has eleven free parameters: the masses and yields of \rhos, \fnm and \ftm, the strength of the interference term for \rhos, and four parameters for the smooth background function $F_{\mathrm{BG}}(M_{\mathrm{\pi\pi}})$). The width of \fnm is limited to be within 40--100~\mvcc~\cite{PDG}. Fits are performed in the mass range $0.45<M_{\pi\pi}<1.7$~\gvcc. The lower limit is selected to include a contribution from \ksnm in the fit but reject contributions from $\eta$, $\eta^{'}(958)$ and \phm mesons, which are difficult to constrain. The upper limit is set to 1.7~\gvcc to account for tails from the \rhos and \ftm contributions. Most of the contributions to the fitting function are well separated in mass, thus reducing the uncertainties of the fit parameters.
 
Examples of the fits in minimum bias pp and 0--20\% and 60--80\% central \pb collisions are shown in Fig.~\ref{fig:analysis:minv} for two different \ptt intervals. The \cn values for the fits are 1.1 (0.9), 0.8 (1.3) and 1.1 (1.2) for pp, 0-20\% and 60-80\% \pb collisions at low(high) transverse momenta, respectively. The contributions of the \ksnm, \omm and \ksm are fixed to the measured particle yields corrected for branching ratios and efficiencies.  The smooth remaining background is described with the function $F_{\mathrm{BG}}(M_{\mathrm{\pi\pi}})$. The remaining contributions from decays of \rhos, \fnm and \ftm mesons are described analytically using the peak model from Section 2.1.3. All fits in different \ptt and centrality intervals result in very reasonable fit probabilities with \cn values close to unity. The yields of \rhos mesons are estimated by integrating the \rhos fitting function in the mass range from $2m_{\mathrm{\pi}}$ to $1.7$~\gvcc. The signal-to-background ratios for \rhos gradually increase with transverse momentum in a range from $10^{-4}$ ($3\times10^{-3}$) to $10^{-2}$ ($7\times10^{-2}$) for 0--20\% (60--80\%) \pb collisions and from $2\times10^{-2}$ to $2\times10^{-1}$ for pp collisions.

\section{Simulations\label{sec:mc}}

Monte-Carlo simulations are used to evaluate the efficiencies for \rhos, \ksnm, \omm and \ksm mesons in the $\pip\pim$ channel as well as to estimate the mass-dependent efficiency corrections for \rhos, \fnm and \ftm. PYTHIA 6~\cite{TSjostrand_JHEP_2006} and PHOJET~\cite{Engel_ZPC_1995,Engel_PRD_1996} were used as event generators for pp collisions, while HIJING~\cite{Wang_PRD_1991} was used to simulate \pb collisions. Signals from the \fnm and \ftm resonances, which are not generated by these codes, were injected  into the simulations. The produced particles and their decay products were traced through the detector materials using GEANT 3~\cite{Brun_CERN_1994}. For each analyzed \ptt and centrality interval, the efficiencies $A\times\effr$ are calculated as the ratio $N_{\mathrm{rec}}/N_{\mathrm{gen}}$, where $N_{\mathrm{rec}}$ is the number of reconstructed particles in the $\pip\pim$ channel after all event and track selection cuts and $N_{\mathrm{gen}}$ is the number of generated mesons within $|y|<0.5$ decaying in the $\rhos, \ksnm, \omm, \fnm, \ftm \rightarrow\pip\pim$, $\omm\rightarrow\pin\pip\pim$ and $\ksm\rightarrow$$K^{\pm}\pi^{\mp}$ channels. In general, the efficiency depends on the shape of the generated particle \ptt spectrum. Therefore, the \ptt spectra of the generated \ksnm, \omm and \ksm mesons are re-weighted to their known or expected shapes. The efficiencies for \rhos are tuned iteratively so that the shapes of the generated \ptt spectra approach the measured shapes. 

\begin{figure}
\begin{center}
\includegraphics[width=25pc]{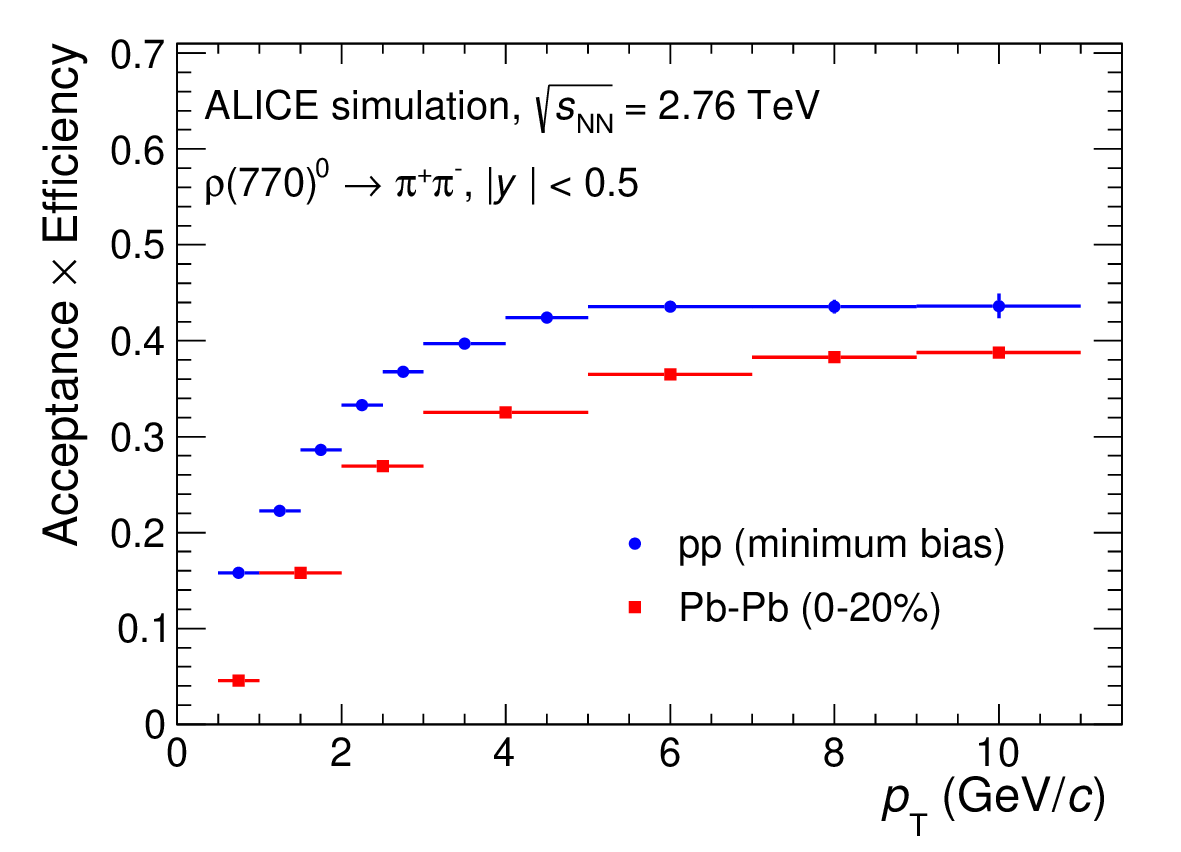}
\end{center}
\caption{\label{fig:mc:receff}(Color online) Acceptance times reconstruction efficiency ($A\times\effr$) evaluated for \rhos meson in pp and central \pb collisions at \rsnn.}
\end{figure}

Examples of efficiencies evaluated for \rhos mesons in pp and the most central \pb collisions as a function of transverse momentum are shown in Fig.~\ref{fig:mc:receff}. The difference in the efficiencies between pp and \pb collisions is expected and is due to the different minimum \ptt cuts and particle identification strategies for daughter particles. In \pb collisions, the efficiencies for \rhos show mild (within 5\%) dependence on collision centrality with a decreasing trend towards more central collisions.

\section{Yield corrections\label{sec:norm}}

In pp collisions the differential transverse momentum spectrum is 

\begin{equation}
\label{eq:norm:invyield}
\frac{\dd^{2}N}{\dd \ptt\dd y} = \frac{1}{N_{\mathrm{MB}}}\frac{\varepsilon_{\mathrm{trig}}}{\varepsilon_{\mathrm{vrtx}}}\frac{1}{A\times\effr}\frac{1}{BR}\frac{N_{\mathrm{\rhos\rightarrow\pi\pi}}}{\Delta y \Delta \ptt},
\end{equation}

where $N_{\mathrm{\rhos\rightarrow\pi\pi}}$ is the \rhos meson yield measured in a given rapidity ($\Delta y$) and transverse momentum ($\Delta\ptt$) interval, $N_{\mathrm{MB}}$ is the number of analyzed minimum bias events, $BR$ and $A\times\effr$ are the resonance branching ratio and efficiency in the $\pip\pim$ decay channel, $\varepsilon_{\mathrm{trig}}= (88.1^{+5.9}_{-3.5})\%$ is a trigger efficiency correction to obtain resonance yields per inelastic pp collision~\cite{Abelev_EPJ_2013} and $\varepsilon_{\mathrm{vrtx}}= 91\pm2\%$ is a vertex cut efficiency correction that accounts for the fraction of \rhos mesons lost after imposing the $z$-vertex cut of 10 cm at the stage of event selection. For the trigger configuration used in this analysis, the number of \rhos mesons in non-triggered events is negligible and no corresponding correction is needed.

For \pb collisions the trigger and vertex cut efficiency corrections, $\varepsilon_{\mathrm{trig}}$ and $\varepsilon_{\mathrm{vrtx}}$, are set to unity. The number of minimum bias events $N_{\mathrm{MB}}$ is replaced with the number of events analyzed in a given centrality interval.  

\section{Systematic uncertainties\label{sec:syst}}

 The total systematic uncertainty is dominated by yield extraction, particle identification and track selection cuts as well as by the global tracking efficiency uncertainties as summarized in Table~\ref{table:systab}. 

\begin{table}[h]
\begin{center}
\begin{tabular}{ r r r }
\hline\hline
Source & \multicolumn{1}{c}{pp} & \pb \\\hline
Yield extraction & 4--13 & 7--13 \\
Particle identification & 4 & 5 \\
Tracking and analysis cuts & 8--9 & 10 \\
Total & 10--16 & 14--17 \\\hline\hline
\end{tabular}
\end{center}
\caption{Relative systematic uncertainties (in \%) for \rhos meson yields in pp and \pb collisions at \rsnn. The single valued uncertainties are \ptt and centrality independent. Values given in ranges correspond to minimum and maximum uncertainties.}
\label{table:systab}
\end{table}

The yield extraction uncertainty is estimated by varying the \rhos meson peak shape, smooth background function, fitting range, temperature parameter in the phase space correction and the relative contributions of \ksnm, \omm and \ksm in the hadronic cocktail. Two peak models, with or without the interference term, are used to extract the \rhos meson parameters from the invariant mass distributions. Fits without the interference term result in lower but still acceptable fit probabilities as well as in systematically lower yields and smaller reconstructed masses. This is the only source of asymmetric systematic uncertainties and it dominates the total uncertainties at low momentum.  The difference in the extracted yields is $\sim10\%$ at low momentum and decreases to $\sim1\%$ for $4-6$~\gvc. For the smooth background function, a fifth order polynomial has been used instead of the $F_{\mathrm{BG}}(M_{\mathrm{\pi\pi}})$ function described in Section 2.1.1. This polynomial has a larger number of fit parameters and could provide an alternative description of the remaining background. The fitting range cannot be varied at its lower edge: it is difficult to control the contributions from $\eta$, $\eta^{'}$(958) and $\phi$(1020) at invariant mass below $0.45$~\gvcc, but it is necessary to account for \ksnm decays resulting in a peak at $0.5$~\gvcc. Instead, the upper limit of the fitting range is varied from $1.7$ to $1.1$~\gvcc, thus excluding the \ftm from the fit and allowing the background function to be more flexible in the narrower fitting range. The temperature parameter in the phase space correction is varied by $\pm25$~MeV to cover the variation of the kinetic freeze-out temperature with multiplicity~\cite{BAbelev_PRC_2013,BAbelev_PLB_2014}. The normalizations of the $\ksnm$, \omm and \ksm templates in the cocktail are independently increased and decreased by the uncertainties of the particle yields and efficiencies estimated to be $\pm30\%$, $\pm20\%$ and $\pm25\%$, respectively. The larger variation for \ksnm is due to the statistical uncertainties of the efficiency, which is only 0.5\% on average. This results in negligible variation of the extracted \rhos meson parameters. The large variation for \omm is dominated by uncertainties in the determination of the $\omega/\pi$ ratio as described in Section 2.1.2. The total yield extraction uncertainty varies from 13(10--13)\% at low momentum to 4(8)\% at intermediate momentum and to 6(7--8)\% at high transverse momentum in pp (\pb) collisions, with rather weak centrality dependence.
   
The particle identification uncertainty is estimated by varying the selection criteria used in analysis. Then in pp collisions the meson yields obtained with ($-1.5\stpc$, $1.5\stpc$) and ($-2.0\stpc$, $1.0\stpc$) particle identification cuts in the TPC are compared to the default value obtained with a $2\stpc$ cut. In \pb collisions, the particle identification cuts are varied to be ($-1.5\stpc$, $1.5\stpc$) and ($-2.0\stpc$, $1.0\stpc$) for tracks that are not matched to the TOF. For tracks with a signal in the TOF, the alternative particle identification cuts are ($-1.5\stof$, $1.5\stof$) and ($-2.0\stof$, $1.0\stof$). In the latter case a variation of the applied $5\stpc$ cut gives a negligible contribution to the systematic uncertainty. As in the case of pp collisions, the meson yields obtained with the varied particle identification cuts are compared to the default value. The resulting uncertainty for the yields is estimated to be 4\% in pp and 5\% in \pb collisions with no centrality dependence.

The global tracking efficiency uncertainty is defined by mismatches between the measured data and Monte Carlo in the probabilities for TPC tracks to be matched to signals in the ITS~\cite{Abelev_PRC_2015,Kishora}. The uncertainty for single tracks is doubled to account for \rhos mesons, which are reconstructed in a decay channel with two charged tracks in the final state. The global tracking uncertainty partially cancels out when ratios of integrated yields, $\rho / \pi$, are calculated. The track selection cuts are varied to estimate the corresponding changes in the fully corrected yields. It is found that the results are sensitive to variation of the number of crossed rows in the TPC and the primary vertex proximity cuts in the transverse plane for reconstructed tracks. The combined systematic uncertainty for the differential yields is estimated to be 8--9 (10)\% in pp (\pb) collisions, with practically no \ptt or centrality dependence.  

Uncertainties in the determination of centrality percentiles result in normalization uncertainties for the measured \rhos yields. The corresponding uncertainties are estimated to be 0.6\%, 1.5\%, 2.95\% and 5.85\% in 0--20\%, 20--40\%, 40--60\% and 60--80\% central \pb collisions using the numbers reported in~\cite{BAbelev_PRC_2013}.

The total systematic uncertainties are calculated as the sum in quadrature of the different contributions and are summarized in Table~\ref{table:systab}.

\section{Results and discussion\label{sec:results}}

\subsection{Particle masses\label{sec:results:masses}}

\begin{figure}
\begin{center}
\includegraphics[width=25pc]{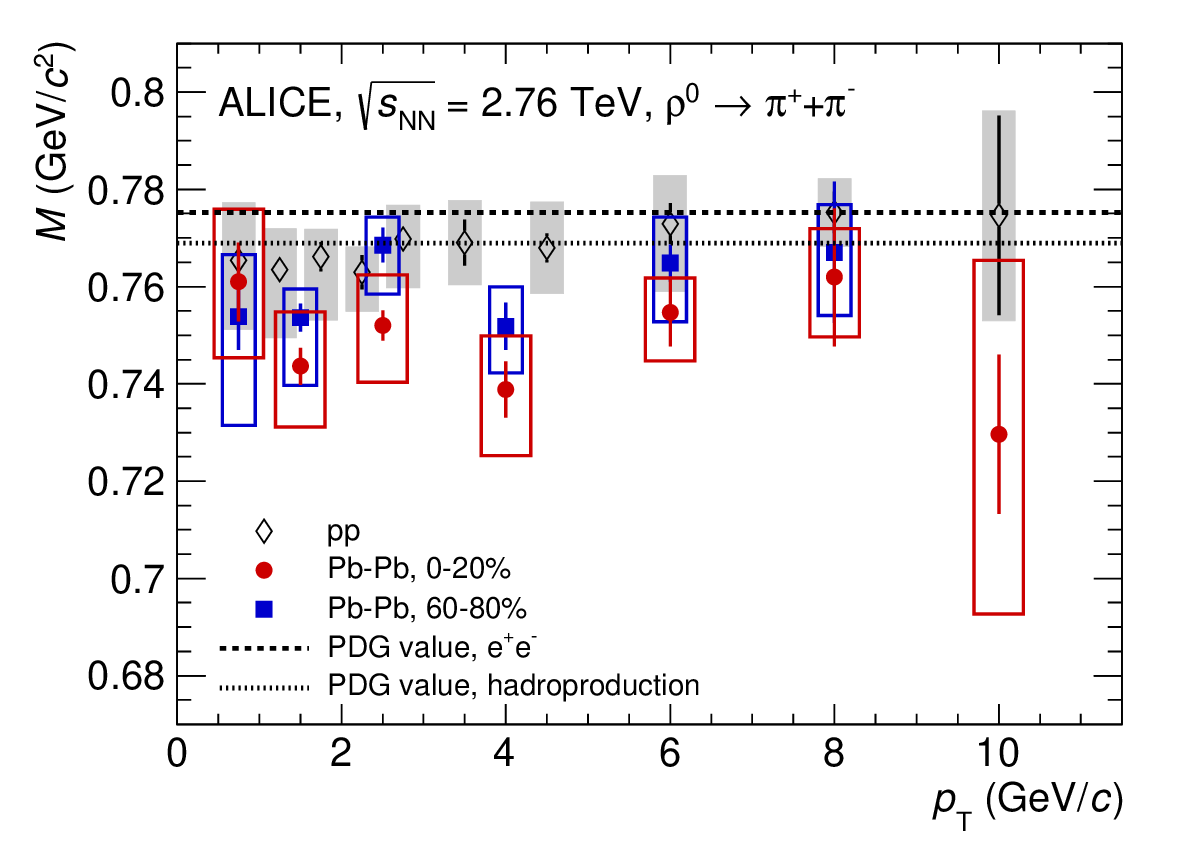}
\caption{\label{fig:results:mass}Reconstructed masses of \rhos mesons as a function of transverse momentum in minimum bias pp, 0--20\% and 60--80\% \pb collisions at \rsnn. The statistical and systematic uncertainties are shown as bars and boxes, respectively. The width of the boxes is varied for visibility. The dashed lines show the \rhos masses as given in ~\cite{PDG}.}
\end{center}
\end{figure}

The dependence of the reconstructed \rhos meson mass on transverse momentum in minimum bias pp,  0--20\% and  60--80\% \pb collisions at \rsnn is shown in Fig.~\ref{fig:results:mass}. The measurements for the 20--40\% and 40--60\% centrality intervals are not shown here, but are similar to the plotted results; these are available in the High Energy Physics Data Repository.
The systematic uncertainties, shown with boxes, account for mass variations from all sources considered in Section 5. The asymmetric part of the systematic uncertainties is from the systematically smaller masses extracted for \rhos mesons using the peak model without the interference term. Two dashed horizontal lines in Fig.~\ref{fig:results:mass} correspond to the \rhos masses quoted in~\cite{PDG} for mesons produced in $e^{+}e^{-}$ annihilation and hadronic interactions. The difference between the values of the masses can be explained by pion scattering as described in ~\cite{Barz:1991fe}. For pp collisions, the reconstructed mass is consistent with the hadroproduced \rhos-meson mass within uncertainties. In \pb collisions, central values of the reconstructed masses are lower by up to $30$~\mvcc with no strong dependence on collision centrality. However, rather large uncertainties prevent any strong conclusions on the mass shift. The STAR data (not shown) also show a tendency for lower masses for \rhos mesons in 40--80\% central \au collisions at \rsnndv~\cite{Adams_PRL_2004_BE}.
 
\subsection{Transverse momentum spectra\label{sec:results:spectra}}

\begin{figure}
\begin{center}
\includegraphics[width=25pc]{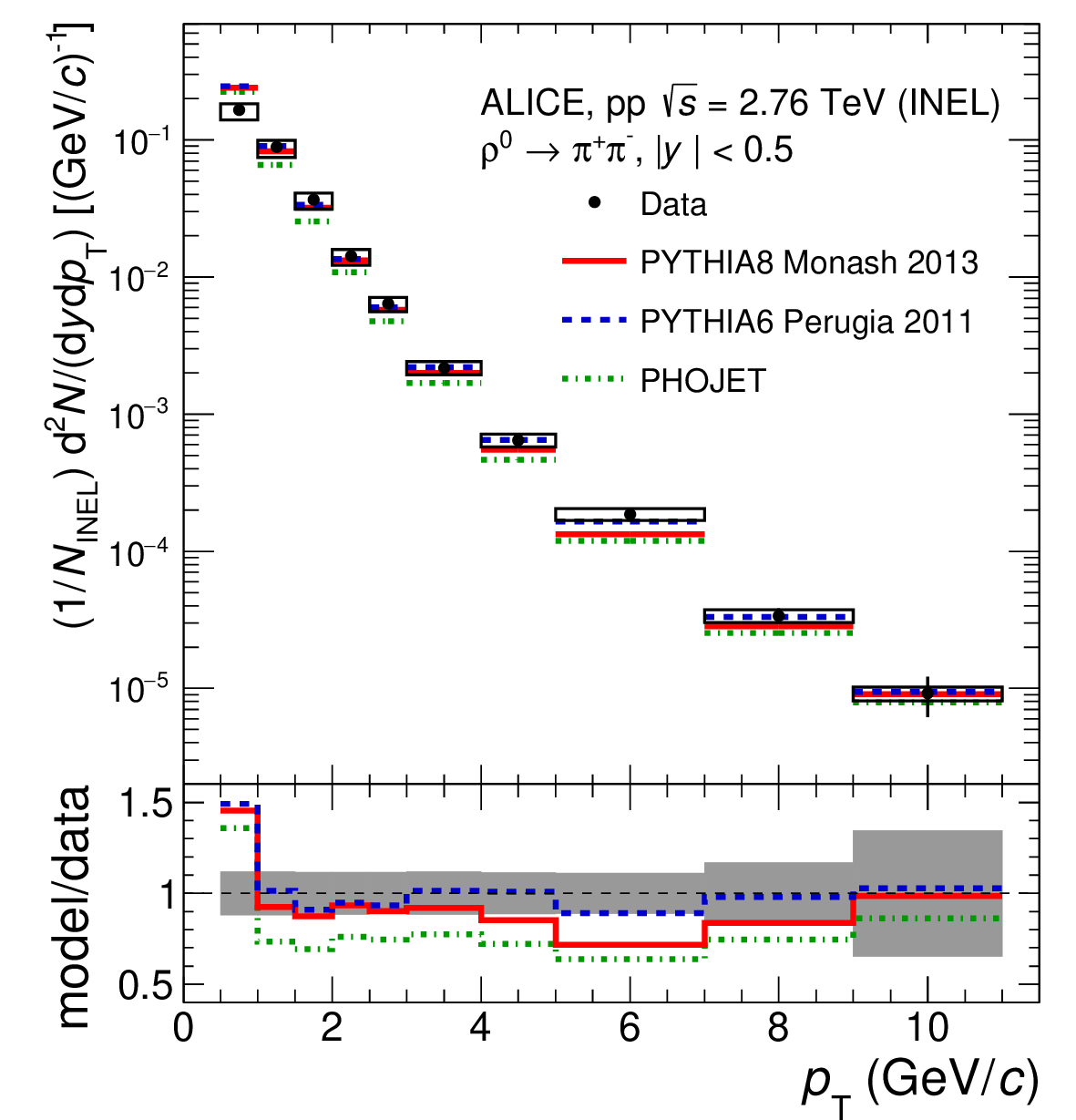}
\end{center}
\caption{\label{fig:results:ptpp}(Color online) Differential yields of \rhos as a function of transverse momentum in inelastic pp collisions at \rs. The statistical and systematic uncertainties are shown as bars and boxes, respectively. The results are compared with model calculations from PYTHIA 6 (Perugia 2011 tune)~\cite{Skands_PRD_2010}, PYTHIA 8.14 (Monash 2013 tune)~\cite{Sjostrand_CPC_2008} and PHOJET~\cite{Engel_ZPC_1995,Engel_PRD_1996}. The lower panel shows the model-to-data ratios and the the gray shaded region represents the sum in quadrature of the systematic and statistical uncertainties associated with the data.}
\end{figure}

\begin{figure}
\begin{center}
\includegraphics[width=25pc]{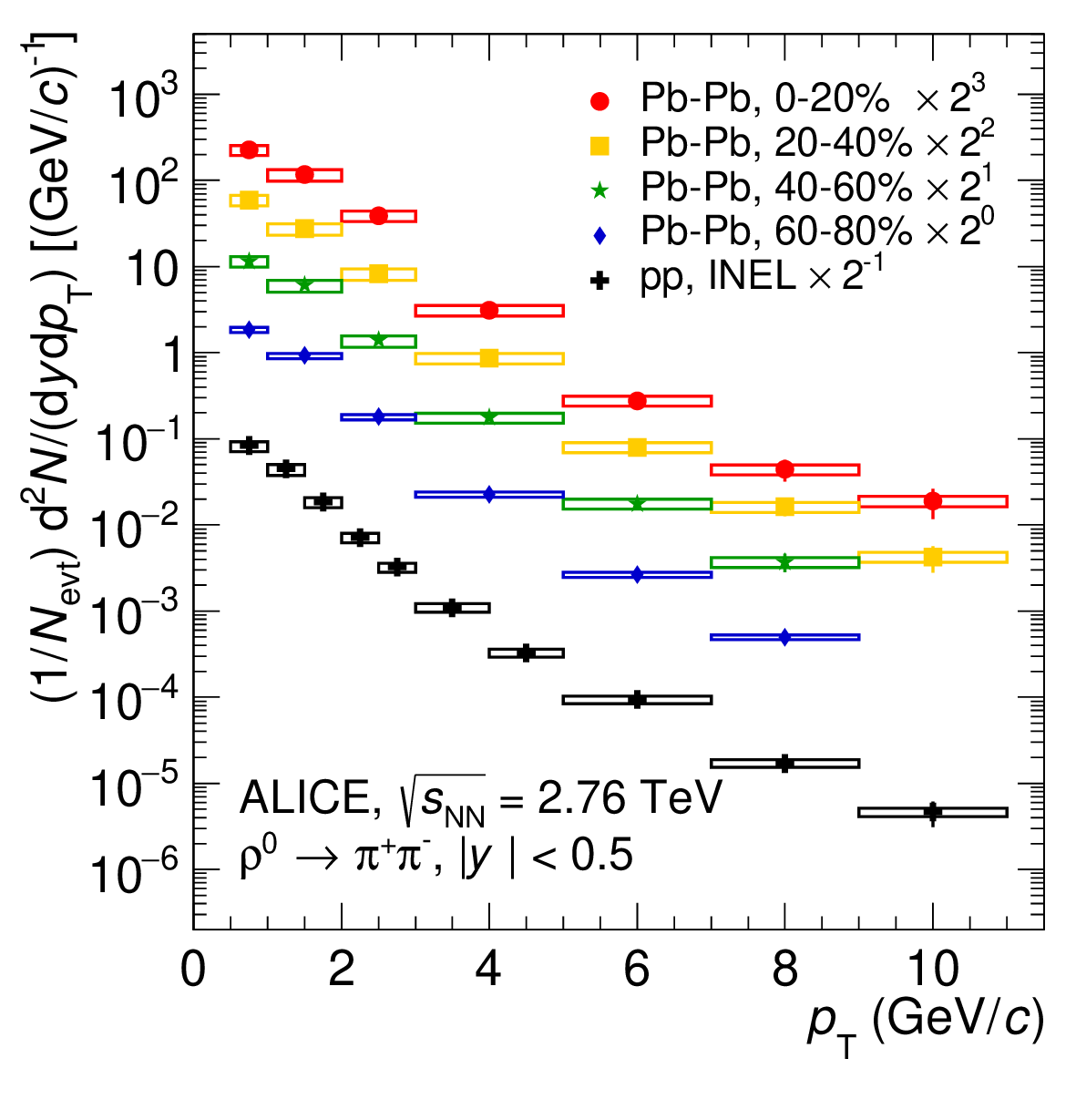}
\end{center}
\caption{\label{fig:results:ptpb} (Color online) Differential yields of \rhos in pp and 0--20\%, 20--40\%, 40--60\%, 60--80\% central \pb collisions at \rsnn. The statistical and systematic uncertainties are shown as bars and boxes, respectively. }
\end{figure}

The differential yields measured for \rhos mesons as a function of transverse momentum in inelastic pp and centrality differential \pb collisions at \rsnn are shown in Figs.~\ref{fig:results:ptpp} and~\ref{fig:results:ptpb}, respectively. The measurements span a wide \ptt range from $0.5$ to $11$~\gvc.

In Fig.~\ref{fig:results:ptpp}, the \ptt spectrum in pp collisions is compared to model calculations from PYTHIA 8.14 (Monash 2013 tune)~\cite{Sjostrand_CPC_2008, Skands:2014pea}, PHOJET~\cite{Engel_ZPC_1995,Engel_PRD_1996} and PYTHIA 6 (Perugia 2011 tune)~\cite{Skands_PRD_2010}. PYTHIA and PHOJET are event generators, which simulate hadronization using the Lund String fragmentation model~\cite{Andersson_PR_1983}. The lower panel of the figure shows the model-to-data ratios as lines and the total uncertainty of the \rhos measurement with a grey band. In general, these models tend to overestimate \rhos-meson production at low momentum, $\ptt<1$~\gvc. PHOJET underestimates \rhos-meson production at intermediate momentum, the best agreement with data is provided by PYTHIA 6 Perugia 2011.

In Fig.~\ref{fig:results:ptpb} the production spectra of \rhos in pp and \pb collisions are shown. The spectra are fit with a L\'{e}vy-Tsallis function~\cite{Tsallis_1988} in the transverse momentum range $\ptt<7$~\gvc to estimate the meson yields outside of the measured range ($\ptt<0.5$~\gvc). The fits are used to calculate the integrated yields (\dndy) and mean transverse momenta (\mpt) following a procedure described in~\cite{Abelev_PRC_2015,Kishora}. The \dndy and \mpt values are evaluated using the data in the measured range and the fit function at lower momentum. The fraction of the total integrated yield in the extrapolated region varies from 30\% in pp collisions to 20 (25)\% in central (peripheral) \pb interactions. Alternative fitting functions, such as Boltzmann-Gibbs blast-wave~\cite{BW_1993}, $m_{\mathrm{T}}$-exponential and power-law functions, are used to fit the measured spectra in different \ptt ranges and evaluate systematic uncertainties for \dndy and \mpt from the extrapolation. The resulting values of \dndy and \mpt are summarized in Table~\ref{table:res} along with their statistical and systematic uncertainties.
 
\begin{table}
\begin{center}
\setlength\extrarowheight{5pt}
\begin{tabular}{ r r r r }
\hline\hline
Collision system & \multicolumn{1}{c}{\dndy} & \multicolumn{1}{c} {\mpt (\gvc)} & \multicolumn{1}{c} {$\rhos / \pi$} \\\hline
Inelastic pp   & \multicolumn{1}{c}{$0.235\pm0.003^{+0.032}_{-0.041}$}               & $0.901\pm0.006^{+0.039}_{-0.045}$  &  $0.126\pm 0.002^{+0.015}_{-0.020}$ \\
\pb, 0--20\%  & $42.90\pm2.59^{+6.04}_{-6.91}\pm0.26$    & $1.191\pm0.031^{+0.095}_{-0.096}$  & $0.076\pm 0.005^{+0.009}_{-0.011}$ \\
\pb, 20--40\% & $21.01\pm0.91^{+2.90}_{-3.40}\pm0.32$    & $1.162\pm0.023^{+0.064}_{-0.067}$  & $0.083\pm 0.004^{+0.009}_{-0.012}$ \\
\pb, 40--60\% & $8.67\pm0.45^{+1.26}_{-1.44}\pm0.26$     & $1.143\pm0.028^{+0.064}_{-0.067}$  & $0.089\pm 0.005^{+0.011}_{-0.013}$ \\
\pb, 60--80\% & $2.74\pm0.13^{+0.41}_{-0.46}\pm0.16$     & $1.083\pm0.024^{+0.070}_{-0.072}$  & $0.101\pm 0.005^{+0.012}_{-0.015}$ \\\hline\hline
\end{tabular}
\end{center}
\caption{Integrated yields (\dndy), mean transverse momenta (\mpt) and $\rhos / \pi$ ratios in pp and centrality differential \pb collisions at \rsnn. For each value the first uncertainty is statistical. For yields the second uncertainty is systematic, but it does not include the normalization uncertainty associated with the centrality selection in \pb collisions. The normalization uncertainty in \pb is reported as the third uncertainty for the yields. For $\rho / \pi$ and \mpt the second uncertainty is the total systematic uncertainty. The asymmetric part of the systematic uncertainties comes from the use of the S\"{o}ding interference term in the fitting function and is correlated between collision systems.   }
\label{table:res}
\end{table}

\subsection{\ptt-integrated particle ratios\label{sec:results:pTint}}

\begin{figure}
\begin{center}
\includegraphics[width=25pc]{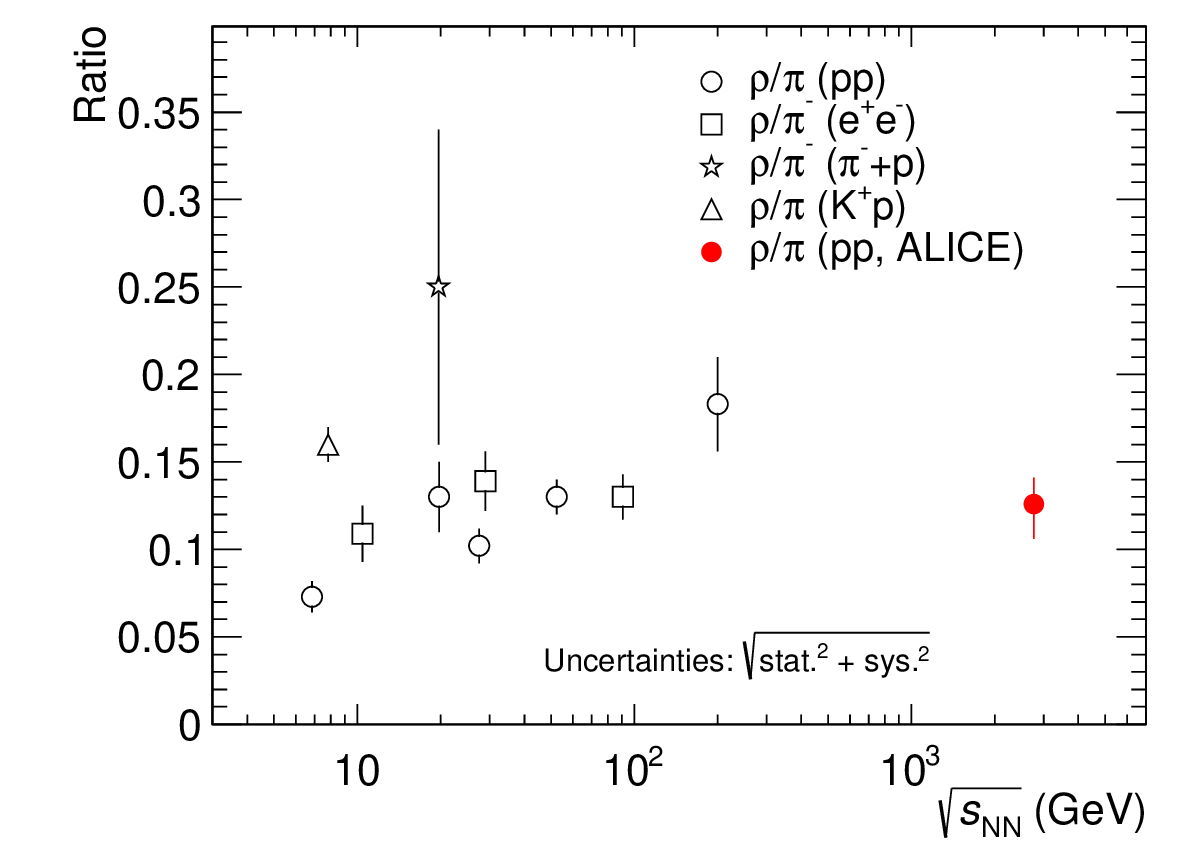}
\end{center}
\caption{\label{fig:results:rhopipp} (Color online) Compilation of $2\rhos/(\pip+\pim)$ and $\rhos/\pim$ measurements at different energies. The ratios are from measurements in $e^{+}e^{-}$ collisions at $\rso=10.45$~GeV~\cite{Albrecht_ZPC_1994}, $29$~GeV~\cite{Derrick_PLB_1985} and $91$~GeV~\cite{Pei_ZPC_1996}; pp collisions at $6.8$~GeV~\cite{Blobel_PLB_1974}, $19.7$~GeV~\cite{Singer_PLB_1976}, $27.5$~GeV~\cite{Aguilar_ZPC_1991}, $52.5$~GeV~\cite{Drijard_ZPC_1981} and $200$~GeV~\cite{Adams_PRL_2004_BE}; $K^{+}$p collisions at $7.82$~GeV~\cite{Chliapnikov_NPB_1980} and $\pim$p collisions at $19.6$~GeV~\cite{Winkelmann_PLB_1975}. The ALICE measurement in pp collisions at \rs is shown with a red marker. The total uncertainties are shown as bars.}
\end{figure}

The collision energy dependence of the $\rhos/\pi$ ratio is presented in Fig.~\ref{fig:results:rhopipp}~\cite{Adams_PRL_2004_BE}. The ALICE result in pp collisions at \rs obtained using charged pion yields from ~\cite{Abelev_PLB_2014} is in good agreement with lower energy measurements and with thermal model predictions for pp collisions at \rsdv~\cite{Becattini_EPJ_2010}, \rs~\cite{Floris:2014pta} and \rssv~\cite{Becattini_JPG_2011}.

\begin{figure}
\includegraphics[width=20pc]{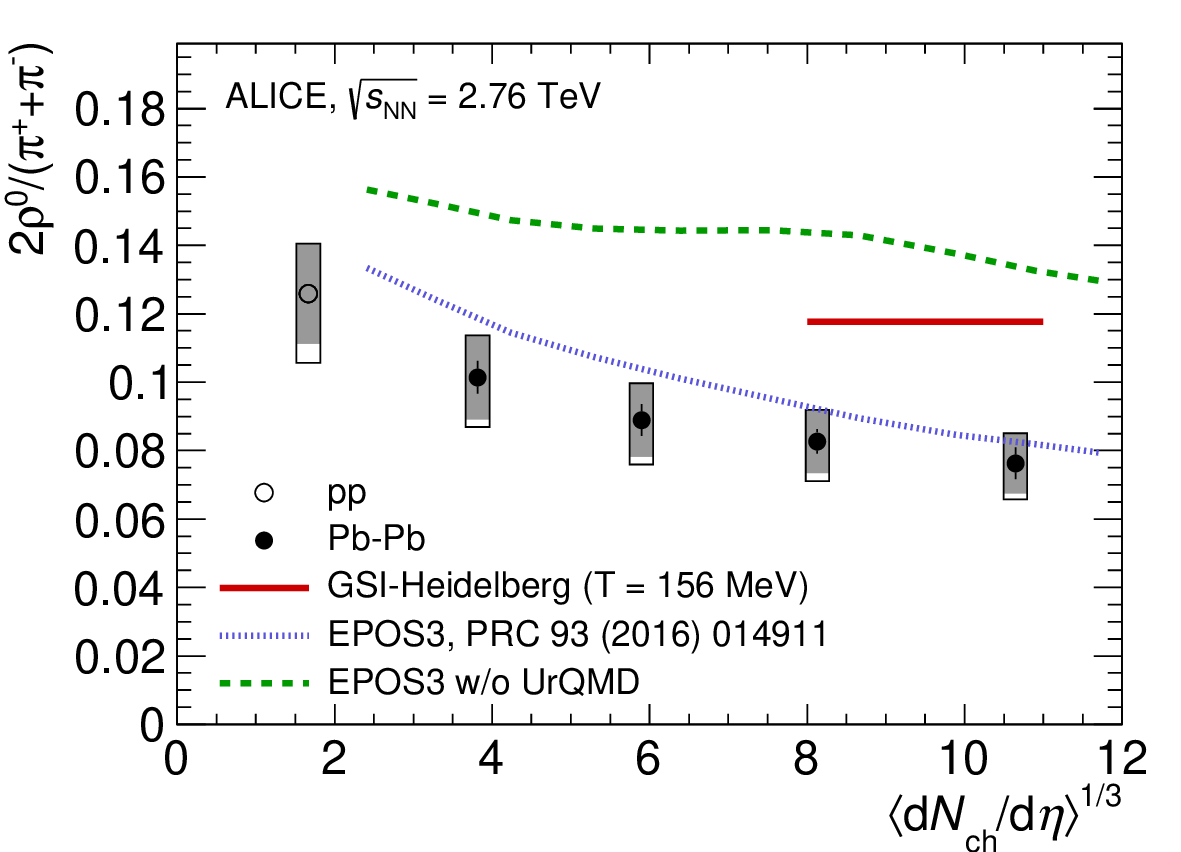}
\includegraphics[width=20pc]{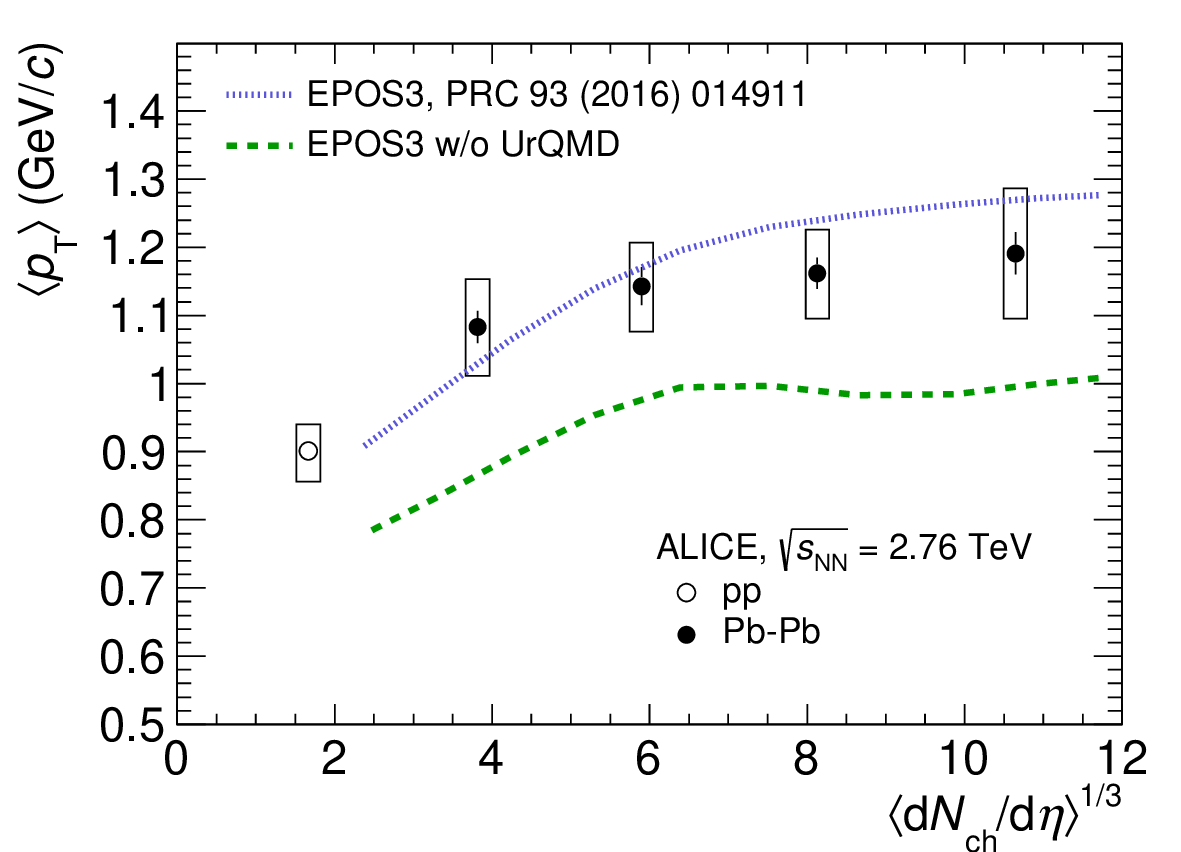}
\caption{\label{fig:results:rhopipb} (Color online) $\rhos/\pi$ ratio (left) and \mpt (right) at mid-rapidity ($|y| < 0.5$) in pp and 0--20\%, 20--40\%, 40--60\%, 60--80\% central \pb collisions at \rsnn as a function of \dncr. Statistical uncertainties are shown as bars. The total and uncorrelated systematic uncertainties are shown with open and shaded boxes, respectively. The widths of the boxes are fixed to arbitrary values for better visibility. The measurements are compared to EPOS3~\cite{Knospe_PRC_2016} calculations. The $\rhos/\pi$ ratio is also compared to grand canonical thermal model~\cite{Thermus} prediction shown with the red horizontal line.}
\end{figure}

The left panel of Fig.~\ref{fig:results:rhopipb} shows the $\rhos/\pi$ ratio measured as a function of \dncr at mid-rapidity ~\cite{Aamodt_PRL_2011} in pp and \pb collisions at \rsnn. In \pb collisions \dncr is used as a proxy for the system size~\cite{Aamodt:2011mr}. The charged pion yields are taken from~\cite{BAbelev_PRC_2013}. The bars represent the statistical uncertainties and the total systematic uncertainties are shown with open boxes. The part of the systematic uncertainties related to the interference term in the \rhos-meson peak model is correlated between points and addition of this term shifts the points in a similar way in pp and \pb collisions. The uncorrelated systematic uncertainties are shown with shaded boxes.

The measured $\rhos/\pi$ ratio in \pb collisions is compared to predictions from a grand-canonical thermal model with a chemical freeze-out temperature of $156$~MeV~\cite{Thermus}. The model is consistent with data only in peripheral collisions. The $\rhos/\pi$ ratio shows a suppression from pp to peripheral \pb and then to central \pb collisions by about 40\%. An analogous suppression was previously observed for short-lived \ksm mesons ($\tau\sim4.2$~fm/$c$) measured in the $\ksm\rightarrow$$K^{\pm}\pi^{\mp}$ decay channel at RHIC and the LHC: the $K^{*0}/K$ ratio was similarly suppressed in central heavy-ion collisions with respect to its value in pp collisions~\cite{Abelev_PRC_2015,Adams_PRC_2005}. The suppression was explained by rescattering of the \ksm daughter particles in the dense hadron gas phase between chemical and kinetic freeze-out. A similar explanation may apply for \rhos mesons, which have a lifetime three times shorter than \ksm and a higher probability to decay before kinetic freeze out.

The measured results are also compared with EPOS3~\cite{Knospe_PRC_2016} calculations. EPOS3 models the evolution of heavy-ion collisions, with initial conditions described by the Gribov-Regge multiple-scattering framework.  The high-density core of the collision is simulated using 3+1 dimensional viscous hydrodynamics and is surrounded by a corona in which decaying strings hadronize.  After the core hadronizes, the evolution of the full system is simulated using UrQMD~\cite{URQMD1,URQMD2}, which includes rescattering and regeneration effects. Calculations were performed with and without a hadronic cascade modeled with UrQMD. Without UrQMD, no significant system size dependence is predicted for the ratio. When UrQMD is enabled, the measured evolution of the $\rhos/\pi$ ratio with multiplicity is well reproduced in \pb collisions (cf. Fig.~\ref{fig:results:rhopipb}, left panel). This suggests that the observed suppression of the \rhos indeed originates from rescattering of its daughter particles in the hadronic phase. EPOS3 was also successful in description of $K^{*0}/K$ ratio in \pb collision~\cite{Kishora,Knospe_PRC_2016}. Under assumption that all suppression for $\rhos$ is from hadronic phase effects,  the same lifetime of the hadronic phase is needed to suppress $K^{*0}$ and $\rhos$.

In the right panel of Fig.~\ref{fig:results:rhopipb} the obtained values of mean \mpt in pp and \pb collisions are reported as a function of the multiplicity. The \mpt values estimated for \rhos by EPOS3 in \pb collisions show an increase as a function of the multiplicity.  The calculation with UrQMD reproduces the measured values in \pb collisions reasonably well, while the calculation without UrQMD significantly underestimates the data.

\subsection{\ptt-differential particle ratios\label{sec:results:pTdiff}}

\begin{figure}[h]
\begin{center}
\includegraphics[width=25pc]{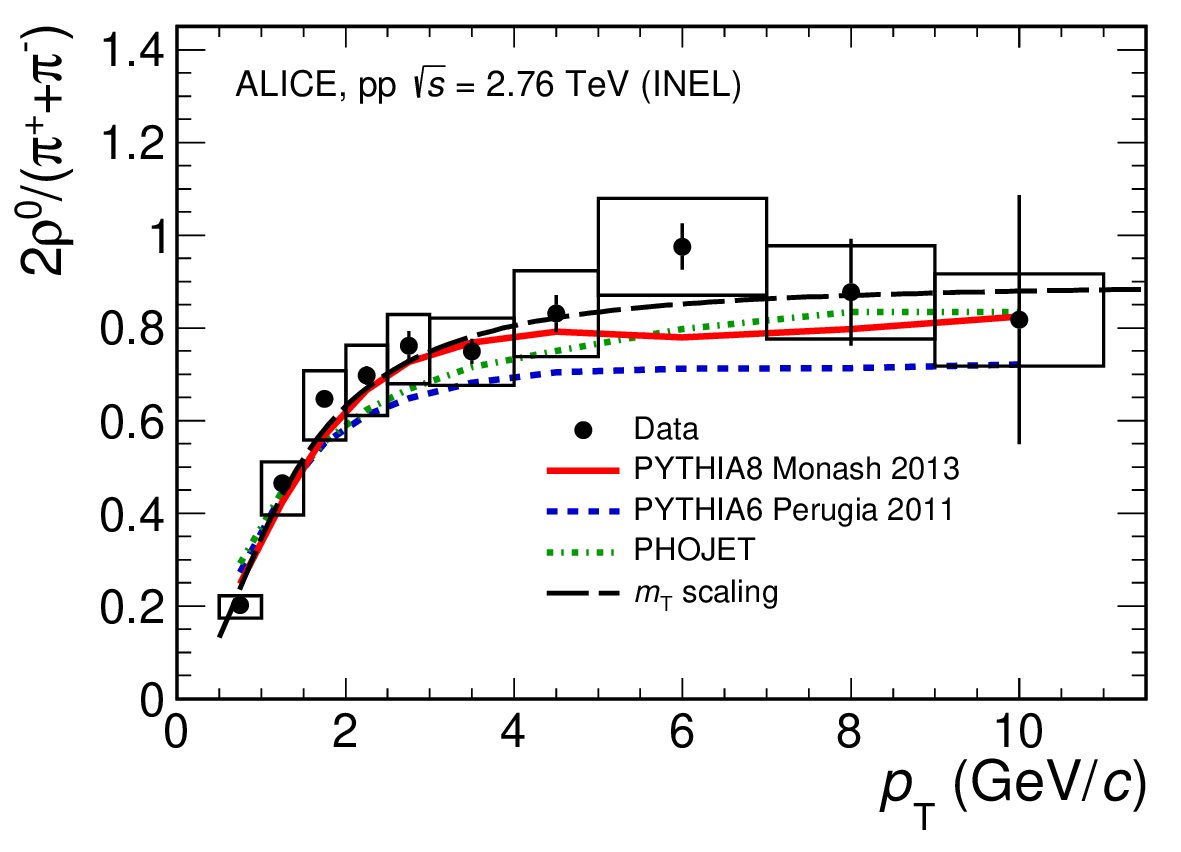}
\end{center}
\caption{\label{fig:results:rhopipppt} (Color online) $\rhos/\pi$ ratio in pp collisions at \rs as a function of transverse momentum. The statistical and systematic uncertainties are shown as bars and boxes, respectively. The results are compared with model calculations from PYTHIA Perugia 2011~\cite{Skands_PRD_2010}, 
PYTHIA 8.14 Monash 2013~\cite{Sjostrand_CPC_2008} and PHOJET~\cite{Engel_ZPC_1995,Engel_PRD_1996}.}
\end{figure}

\begin{figure}
\includegraphics[width=38pc]{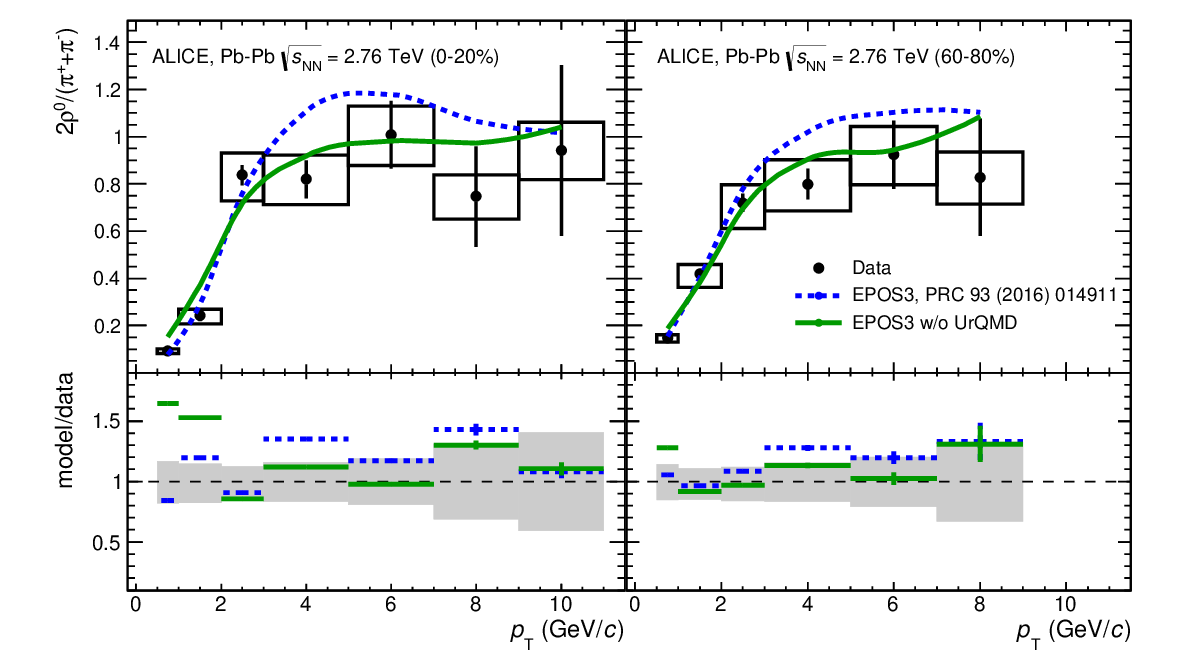}
\caption{\label{fig:results:rhopipbpt} (Color online) $\rhos/\pi$ ratio in 0--20\% (left panel) and 60--80\% (right panel) central \pb collisions at \rsnn. The statistical and systematic uncertainties are shown as bars and boxes, respectively. The measurements are compared to EPOS3~\cite{Knospe_PRC_2016} calculations performed with and without a hadronic cascade modeled with UrQMD~\cite{URQMD1,URQMD2}. In the lower panels, the model to data ratio is reported. Bars indicate statistical uncertainty of calculations whereas the gray shaded boxes represent the square root of sum of squares of statistical and systematic uncertainties associated with data.}
\end{figure}

The $\rhos/\pi$ ratios measured in pp and \pb collisions (in the 0--20\%, 60--80\% centrality intervals) at \rsnn as a function of transverse momentum are shown in Figs.~\ref{fig:results:rhopipppt} and~\ref{fig:results:rhopipbpt}, respectively. The \ptt spectra for pions are obtained from~\cite{BAbelev_PRC_2013,Adam_PRC_2016}.

The ratio in pp collisions is compared to the same model calculations as in Fig.~\ref{fig:results:ptpp}.  As for the \ptt spectra, the models overestimate $\rhos/\pi$ ratio at low momenta, $\ptt<1$~\gvc. At higher momentum, the predictions of the event generators differ by tens of percent, with PYTHIA 8.14 Monash 2013~\cite{Sjostrand_CPC_2008} and PHOJET~\cite{Engel_ZPC_1995,Engel_PRD_1996} providing the best description of the data. The $m_{\mathrm{T}}$-scaling curve shown in the figure is obtained in the same way as in Fig.~\ref{fig:analysis:ompi}. It is normalized to $\rhos/\pi = 0.88$ at high momentum, which is obtained from the fit to data points at $\ptt>4$~\gvc. The curve very well reproduces the measurement results in the whole range of measurements. We also note that the $\rhos/\pi$ ratio measured in pp collisions at \rs is very close to the $\omega/\pi$ ratio measured at lower energies and presented in Fig.~\ref{fig:analysis:ompi}. This is consistent with PYTHIA, which predicts very weak energy dependence of the $\rhos/\pi$ and $\omega/\pi$ ratios, with $\rhos/\omega\sim1.05$ in the measured \ptt range.
 
The $\rhos/\pi$ ratio measured in peripheral \pb collisions is very similar to that in pp collisions, as can be seen by comparing Fig.~\ref{fig:results:rhopipppt} and the right panel of Fig.~\ref{fig:results:rhopipbpt}. However, in central \pb collisions the ratio is significantly suppressed at low momentum ($\ptt<2$~\gvc). This means that the suppression of the \ptt-integrated $\rhos/\pi$ ratio reported earlier is due to the suppression of low-\ptt particle production in central \pb collisions. It is important to note that the \ptt-dependent suppression of the $\rhos/\pi$ ratio is reproduced by EPOS3 calculations when the hadronic cascade simulated with UrQMD is taken into account. For $\ptt<2$~\gvc, EPOS3 without UrQMD overestimates the ratio by 30--40\%. This may serve as another indication that \rhos-meson suppression is due to daughter particle rescattering in the hadronic phase.
 	 
\subsection{Nuclear modification factors\label{sec:results:raa}}

The nuclear modification factor \raa is used to study medium-induced effects in heavy-ion collisions.  The \raa is the ratio of the yield of a particle in nucleus-nucleus collisions to its yield in pp collisions.  This ratio is scaled by the number of binary nucleon-nucleon collisions in each centrality class, which is estimated from Glauber model calculations~\cite{Alver:2008aq,Loizides:2014vua}.  For each \ptt bin,

\begin{equation}
\raa=\frac{1}{\langle N_{\mathrm{coll}}\rangle}\cdot\frac{\mathrm{d}N_{\mathrm{AA}}/\mathrm{d}\ptt}{\mathrm{d}N_{\mathrm{pp}}/\mathrm{d}\ptt}.
\end{equation}

\begin{figure}
\includegraphics[width=38pc]{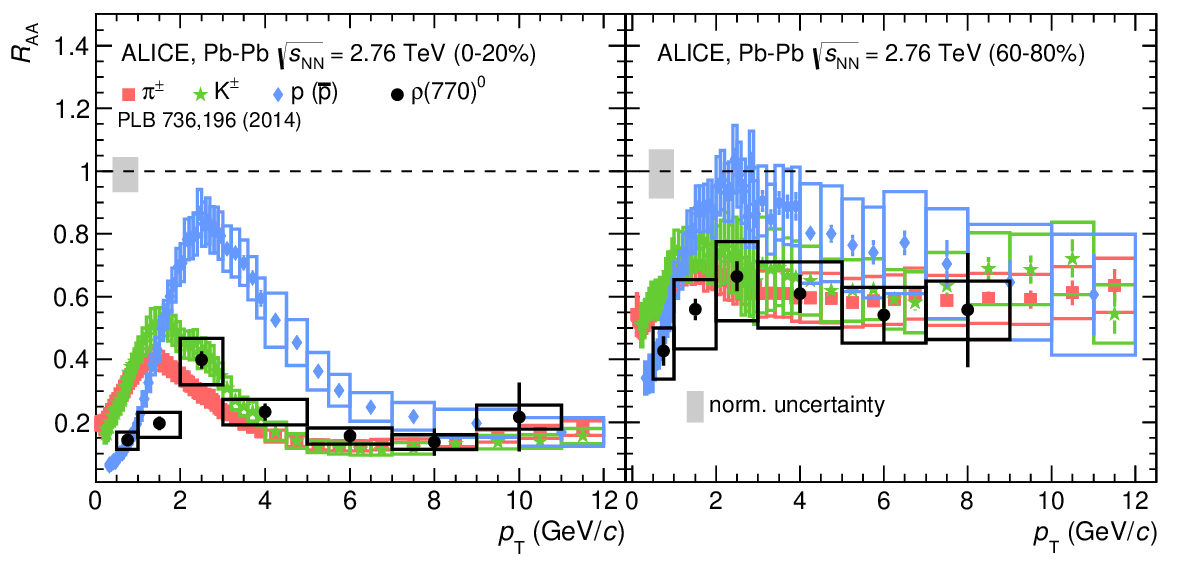}
\caption{\label{fig:results:raa} (Color online) Nuclear modification factors \raa measured for charged pions, charged kaons, (anti)protons~\cite{Adam_PRC_2016} and \rhos mesons in 0--20\% (left panel) and 60--80\% (right panel) central \pb collisions at \rsnn. The statistical and systematic uncertainties are shown as bars and boxes, respectively. The boxes at unity correspond to scaling uncertainties.}
\end{figure}

The nuclear modification factors measured in 0--20\% and 60--80\% central \pb collisions 
at \rsnn for charged pions, charged kaons, (anti)protons~\cite{Adam_PRC_2016} and \rhos mesons are reported in Fig.~\ref{fig:results:raa}. (\raa values for the other centrality classes are available in the High Energy Physics Data Repository.) One can see that in \pb collisions, production of all hadrons is suppressed by a similar amount at high transverse momenta of $\ptt > 8~\gvc$ and there is no dependence of the suppression on particle mass or quark content within uncertainties. This observation, also confirmed by measurements for \ksm and \phm mesons~\cite{Kishora}, rules out models that predict a species-dependent suppression of light hadrons and puts additional constraints on parton energy loss and fragmentation models~\cite{Bellwied:2010pr,Liu:2008zb,Liu:2006sf}.  There is a clear species dependence of \raa at intermediate transverse momentum, which is likely to be a result of an interplay between different effects such as radial flow, low-\ptt suppression, species dependent \ptt shapes of the pp reference spectra~\cite{Abelev_PRC_2015,Kishora}.
 	 
\section{Conclusions\label{sec:conclusions}}
We have measured the production of \rhos mesons in minimum bias pp and centrality differential \pb collisions at \rsnn. These measurements were performed in a wide \ptt range from $0.5$ to $11$~\gvc in the hadronic decay channel $\rhos\rightarrow\pip\pim$. The invariant mass distributions are well described by a hadronic cocktail, assuming the same \rhos-meson peak shape in pp and central heavy-ion collisions. However, alternative peak models for \rhos mesons are not ruled out by data. The reconstructed masses of \rhos mesons are consistent with the hadroproduced mass of \rhos within uncertainties. In pp collisions, the transverse momentum spectrum for $\ptt>1$~\gvc is rather well reproduced by PYTHIA 8.14 Monash 2013 and PYTHIA 6 Perugia 2011, while a better agreement is observed for PYTHIA 8.14 Monash 2013 and PHOJET for the \ptt-differential $\rhos/\pi$ ratio. In \pb collisions, the measured yields for the \rhos meson are suppressed at low momentum ($\ptt<2$~\gvc). The ratio of integrated yields, $\rhos/\pi$, decreases by $\sim40\%$ from pp to central \pb collisions, similar to what was previously observed for the \kskm ratio and explained by rescattering of the daughter particles in the hadron-gas phase. The relative suppression of the \ptt-integrated and \ptt-differential $\rhos/\pi$ ratios is well reproduced by EPOS3 calculations, provided that the hadronic cascade is modeled with UrQMD. This suggests that the observed centrality-dependent suppression of \rhos production occurs due to rescattering of daughter pions in the hadronic phase between chemical and kinetic freeze-out, with the rescattering being most important at low \ptt. However, suppression of \rhos-meson production may also occur due to significant line shape modifications not accounted for in the peak model used in this analysis. The development of a realistic model of the \rhos-meson peak shape in heavy-ion collisions would be an important subject for theoretical studies. Once available, the model predictions can be compared to the 'vacuum shape' results for \rhos reported in this paper and implications for the measured yields due to possible line shape modifications can be inferred and discussed.

\newenvironment{acknowledgement}{\relax}{\relax}
\begin{acknowledgement}
\section*{Acknowledgements}

The ALICE Collaboration would like to thank all its engineers and technicians for their invaluable contributions to the construction of the experiment and the CERN accelerator teams for the outstanding performance of the LHC complex.
The ALICE Collaboration gratefully acknowledges the resources and support provided by all Grid centres and the Worldwide LHC Computing Grid (WLCG) collaboration.
The ALICE Collaboration acknowledges the following funding agencies for their support in building and running the ALICE detector:
A. I. Alikhanyan National Science Laboratory (Yerevan Physics Institute) Foundation (ANSL), State Committee of Science and World Federation of Scientists (WFS), Armenia;
Austrian Academy of Sciences and Nationalstiftung f\"{u}r Forschung, Technologie und Entwicklung, Austria;
Ministry of Communications and High Technologies, National Nuclear Research Center, Azerbaijan;
Conselho Nacional de Desenvolvimento Cient\'{\i}fico e Tecnol\'{o}gico (CNPq), Universidade Federal do Rio Grande do Sul (UFRGS), Financiadora de Estudos e Projetos (Finep) and Funda\c{c}\~{a}o de Amparo \`{a} Pesquisa do Estado de S\~{a}o Paulo (FAPESP), Brazil;
Ministry of Science \& Technology of China (MSTC), National Natural Science Foundation of China (NSFC) and Ministry of Education of China (MOEC) , China;
Ministry of Science and Education, Croatia;
Ministry of Education, Youth and Sports of the Czech Republic, Czech Republic;
The Danish Council for Independent Research | Natural Sciences, the Carlsberg Foundation and Danish National Research Foundation (DNRF), Denmark;
Helsinki Institute of Physics (HIP), Finland;
Commissariat \`{a} l'Energie Atomique (CEA) and Institut National de Physique Nucl\'{e}aire et de Physique des Particules (IN2P3) and Centre National de la Recherche Scientifique (CNRS), France;
Bundesministerium f\"{u}r Bildung, Wissenschaft, Forschung und Technologie (BMBF) and GSI Helmholtzzentrum f\"{u}r Schwerionenforschung GmbH, Germany;
General Secretariat for Research and Technology, Ministry of Education, Research and Religions, Greece;
National Research, Development and Innovation Office, Hungary;
Department of Atomic Energy Government of India (DAE), Department of Science and Technology, Government of India (DST), University Grants Commission, Government of India (UGC) and Council of Scientific and Industrial Research (CSIR), India;
Indonesian Institute of Science, Indonesia;
Centro Fermi - Museo Storico della Fisica e Centro Studi e Ricerche Enrico Fermi and Istituto Nazionale di Fisica Nucleare (INFN), Italy;
Institute for Innovative Science and Technology , Nagasaki Institute of Applied Science (IIST), Japan Society for the Promotion of Science (JSPS) KAKENHI and Japanese Ministry of Education, Culture, Sports, Science and Technology (MEXT), Japan;
Consejo Nacional de Ciencia (CONACYT) y Tecnolog\'{i}a, through Fondo de Cooperaci\'{o}n Internacional en Ciencia y Tecnolog\'{i}a (FONCICYT) and Direcci\'{o}n General de Asuntos del Personal Academico (DGAPA), Mexico;
Nederlandse Organisatie voor Wetenschappelijk Onderzoek (NWO), Netherlands;
The Research Council of Norway, Norway;
Commission on Science and Technology for Sustainable Development in the South (COMSATS), Pakistan;
Pontificia Universidad Cat\'{o}lica del Per\'{u}, Peru;
Ministry of Science and Higher Education and National Science Centre, Poland;
Korea Institute of Science and Technology Information and National Research Foundation of Korea (NRF), Republic of Korea;
Ministry of Education and Scientific Research, Institute of Atomic Physics and Romanian National Agency for Science, Technology and Innovation, Romania;
Joint Institute for Nuclear Research (JINR), Ministry of Education and Science of the Russian Federation and National Research Centre Kurchatov Institute, Russia;
Ministry of Education, Science, Research and Sport of the Slovak Republic, Slovakia;
National Research Foundation of South Africa, South Africa;
Centro de Aplicaciones Tecnol\'{o}gicas y Desarrollo Nuclear (CEADEN), Cubaenerg\'{\i}a, Cuba and Centro de Investigaciones Energ\'{e}ticas, Medioambientales y Tecnol\'{o}gicas (CIEMAT), Spain;
Swedish Research Council (VR) and Knut \& Alice Wallenberg Foundation (KAW), Sweden;
European Organization for Nuclear Research, Switzerland;
National Science and Technology Development Agency (NSDTA), Suranaree University of Technology (SUT) and Office of the Higher Education Commission under NRU project of Thailand, Thailand;
Turkish Atomic Energy Agency (TAEK), Turkey;
National Academy of  Sciences of Ukraine, Ukraine;
Science and Technology Facilities Council (STFC), United Kingdom;
National Science Foundation of the United States of America (NSF) and United States Department of Energy, Office of Nuclear Physics (DOE NP), United States of America.    
\end{acknowledgement}

\bibliography{mybib}{}
\bibliographystyle{utphys}   

\newpage
\appendix
\section{The ALICE Collaboration}
\label{app:collab}

\begingroup
\small
\begin{flushleft}
S.~Acharya\Irefn{org138}\And 
F.T.-.~Acosta\Irefn{org22}\And 
D.~Adamov\'{a}\Irefn{org94}\And 
J.~Adolfsson\Irefn{org81}\And 
M.M.~Aggarwal\Irefn{org98}\And 
G.~Aglieri Rinella\Irefn{org36}\And 
M.~Agnello\Irefn{org33}\And 
N.~Agrawal\Irefn{org49}\And 
Z.~Ahammed\Irefn{org138}\And 
S.U.~Ahn\Irefn{org77}\And 
S.~Aiola\Irefn{org143}
% ...
\And
\renewcommand\labelenumi{\textsuperscript{\theenumi}~}

\section*{Affiliation notes}
\renewcommand\theenumi{\roman{enumi}}
\begin{Authlist}
\item \Adef{orgI}Dipartimento DET del Politecnico di Torino, Turin, Italy
\item \Adef{orgII}M.V. Lomonosov Moscow State University, D.V. Skobeltsyn Institute of Nuclear, Physics, Moscow, Russia
\item \Adef{orgIII}Department of Applied Physics, Aligarh Muslim University, Aligarh, India
\item \Adef{orgIV}Institute of Theoretical Physics, University of Wroclaw, Poland
\end{Authlist}

\section*{Collaboration Institutes}
\renewcommand\theenumi{\arabic{enumi}~}
\begin{Authlist}
\item \Idef{org1}A.I. Alikhanyan National Science Laboratory (Yerevan Physics Institute) Foundation, Yerevan, Armenia
\item \Idef{org2}Benem\'{e}rita Universidad Aut\'{o}noma de Puebla, Puebla, Mexico
\item \Idef{org3}Bogolyubov Institute for Theoretical Physics, National Academy of Sciences of Ukraine, Kiev, Ukraine
\item \Idef{org4}Bose Institute, Department of Physics  and Centre for Astroparticle Physics and Space Science (CAPSS), Kolkata, India
\item \Idef{org5}Budker Institute for Nuclear Physics, Novosibirsk, Russia
\item \Idef{org6}California Polytechnic State University, San Luis Obispo, California, United States
\item \Idef{org7}Central China Normal University, Wuhan, China
\item \Idef{org8}Centre de Calcul de l'IN2P3, Villeurbanne, Lyon, France
\item \Idef{org9}Centro de Aplicaciones Tecnol\'{o}gicas y Desarrollo Nuclear (CEADEN), Havana, Cuba
\item \Idef{org10}Centro de Investigaci\'{o}n y de Estudios Avanzados (CINVESTAV), Mexico City and M\'{e}rida, Mexico
\item \Idef{org11}Centro Fermi - Museo Storico della Fisica e Centro Studi e Ricerche ``Enrico Fermi', Rome, Italy
\item \Idef{org12}Chicago State University, Chicago, Illinois, United States
\item \Idef{org13}China Institute of Atomic Energy, Beijing, China
\item \Idef{org14}Chonbuk National University, Jeonju, Republic of Korea
\item \Idef{org15}Comenius University Bratislava, Faculty of Mathematics, Physics and Informatics, Bratislava, Slovakia
\item \Idef{org16}COMSATS Institute of Information Technology (CIIT), Islamabad, Pakistan
\item \Idef{org17}Creighton University, Omaha, Nebraska, United States
\item \Idef{org18}Department of Physics, Aligarh Muslim University, Aligarh, India
\item \Idef{org19}Department of Physics, Ohio State University, Columbus, Ohio, United States
\item \Idef{org20}Department of Physics, Pusan National University, Pusan, Republic of Korea
\item \Idef{org21}Department of Physics, Sejong University, Seoul, Republic of Korea
\item \Idef{org22}Department of Physics, University of California, Berkeley, California, United States
\item \Idef{org23}Department of Physics, University of Oslo, Oslo, Norway
\item \Idef{org24}Department of Physics and Technology, University of Bergen, Bergen, Norway
\item \Idef{org25}Dipartimento di Fisica dell'Universit\`{a} 'La Sapienza' and Sezione INFN, Rome, Italy
\item \Idef{org26}Dipartimento di Fisica dell'Universit\`{a} and Sezione INFN, Cagliari, Italy
\item \Idef{org27}Dipartimento di Fisica dell'Universit\`{a} and Sezione INFN, Trieste, Italy
\item \Idef{org28}Dipartimento di Fisica dell'Universit\`{a} and Sezione INFN, Turin, Italy
\item \Idef{org29}Dipartimento di Fisica e Astronomia dell'Universit\`{a} and Sezione INFN, Bologna, Italy
\item \Idef{org30}Dipartimento di Fisica e Astronomia dell'Universit\`{a} and Sezione INFN, Catania, Italy
\item \Idef{org31}Dipartimento di Fisica e Astronomia dell'Universit\`{a} and Sezione INFN, Padova, Italy
\item \Idef{org32}Dipartimento di Fisica `E.R.~Caianiello' dell'Universit\`{a} and Gruppo Collegato INFN, Salerno, Italy
\item \Idef{org33}Dipartimento DISAT del Politecnico and Sezione INFN, Turin, Italy
\item \Idef{org34}Dipartimento di Scienze e Innovazione Tecnologica dell'Universit\`{a} del Piemonte Orientale and INFN Sezione di Torino, Alessandria, Italy
\item \Idef{org35}Dipartimento Interateneo di Fisica `M.~Merlin' and Sezione INFN, Bari, Italy
\item \Idef{org36}European Organization for Nuclear Research (CERN), Geneva, Switzerland
\item \Idef{org37}Faculty of Electrical Engineering, Mechanical Engineering and Naval Architecture, University of Split, Split, Croatia
\item \Idef{org38}Faculty of Engineering and Science, Western Norway University of Applied Sciences, Bergen, Norway
\item \Idef{org39}Faculty of Nuclear Sciences and Physical Engineering, Czech Technical University in Prague, Prague, Czech Republic
\item \Idef{org40}Faculty of Science, P.J.~\v{S}af\'{a}rik University, Ko\v{s}ice, Slovakia
\item \Idef{org41}Frankfurt Institute for Advanced Studies, Johann Wolfgang Goethe-Universit\"{a}t Frankfurt, Frankfurt, Germany
\item \Idef{org42}Gangneung-Wonju National University, Gangneung, Republic of Korea
\item \Idef{org43}Gauhati University, Department of Physics, Guwahati, India
\item \Idef{org44}Helmholtz-Institut f\"{u}r Strahlen- und Kernphysik, Rheinische Friedrich-Wilhelms-Universit\"{a}t Bonn, Bonn, Germany
\item \Idef{org45}Helsinki Institute of Physics (HIP), Helsinki, Finland
\item \Idef{org46}Hiroshima University, Hiroshima, Japan
\item \Idef{org47}Hochschule Worms, Zentrum  f\"{u}r Technologietransfer und Telekommunikation (ZTT), Worms, Germany
\item \Idef{org48}Horia Hulubei National Institute of Physics and Nuclear Engineering, Bucharest, Romania
\item \Idef{org49}Indian Institute of Technology Bombay (IIT), Mumbai, India
\item \Idef{org50}Indian Institute of Technology Indore, Indore, India
\item \Idef{org51}Indonesian Institute of Sciences, Jakarta, Indonesia
\item \Idef{org52}INFN, Laboratori Nazionali di Frascati, Frascati, Italy
\item \Idef{org53}INFN, Sezione di Bari, Bari, Italy
\item \Idef{org54}INFN, Sezione di Bologna, Bologna, Italy
\item \Idef{org55}INFN, Sezione di Cagliari, Cagliari, Italy
\item \Idef{org56}INFN, Sezione di Catania, Catania, Italy
\item \Idef{org57}INFN, Sezione di Padova, Padova, Italy
\item \Idef{org58}INFN, Sezione di Roma, Rome, Italy
\item \Idef{org59}INFN, Sezione di Torino, Turin, Italy
\item \Idef{org60}INFN, Sezione di Trieste, Trieste, Italy
\item \Idef{org61}Inha University, Incheon, Republic of Korea
\item \Idef{org62}Institut de Physique Nucl\'{e}aire d'Orsay (IPNO), Institut National de Physique Nucl\'{e}aire et de Physique des Particules (IN2P3/CNRS), Universit\'{e} de Paris-Sud, Universit\'{e} Paris-Saclay, Orsay, France
\item \Idef{org63}Institute for Nuclear Research, Academy of Sciences, Moscow, Russia
\item \Idef{org64}Institute for Subatomic Physics, Utrecht University/Nikhef, Utrecht, Netherlands
\item \Idef{org65}Institute for Theoretical and Experimental Physics, Moscow, Russia
\item \Idef{org66}Institute of Experimental Physics, Slovak Academy of Sciences, Ko\v{s}ice, Slovakia
\item \Idef{org67}Institute of Physics, Bhubaneswar, India
\item \Idef{org68}Institute of Physics of the Czech Academy of Sciences, Prague, Czech Republic
\item \Idef{org69}Institute of Space Science (ISS), Bucharest, Romania
\item \Idef{org70}Institut f\"{u}r Kernphysik, Johann Wolfgang Goethe-Universit\"{a}t Frankfurt, Frankfurt, Germany
\item \Idef{org71}Instituto de Ciencias Nucleares, Universidad Nacional Aut\'{o}noma de M\'{e}xico, Mexico City, Mexico
\item \Idef{org72}Instituto de F\'{i}sica, Universidade Federal do Rio Grande do Sul (UFRGS), Porto Alegre, Brazil
\item \Idef{org73}Instituto de F\'{\i}sica, Universidad Nacional Aut\'{o}noma de M\'{e}xico, Mexico City, Mexico
\item \Idef{org74}iThemba LABS, National Research Foundation, Somerset West, South Africa
\item \Idef{org75}Johann-Wolfgang-Goethe Universit\"{a}t Frankfurt Institut f\"{u}r Informatik, Fachbereich Informatik und Mathematik, Frankfurt, Germany
\item \Idef{org76}Joint Institute for Nuclear Research (JINR), Dubna, Russia
\item \Idef{org77}Korea Institute of Science and Technology Information, Daejeon, Republic of Korea
\item \Idef{org78}KTO Karatay University, Konya, Turkey
\item \Idef{org79}Laboratoire de Physique Subatomique et de Cosmologie, Universit\'{e} Grenoble-Alpes, CNRS-IN2P3, Grenoble, France
\item \Idef{org80}Lawrence Berkeley National Laboratory, Berkeley, California, United States
\item \Idef{org81}Lund University Department of Physics, Division of Particle Physics, Lund, Sweden
\item \Idef{org82}Nagasaki Institute of Applied Science, Nagasaki, Japan
\item \Idef{org83}Nara Women{'}s University (NWU), Nara, Japan
\item \Idef{org84}National and Kapodistrian University of Athens, School of Science, Department of Physics , Athens, Greece
\item \Idef{org85}National Centre for Nuclear Research, Warsaw, Poland
\item \Idef{org86}National Institute of Science Education and Research, HBNI, Jatni, India
\item \Idef{org87}National Nuclear Research Center, Baku, Azerbaijan
\item \Idef{org88}National Research Centre Kurchatov Institute, Moscow, Russia
\item \Idef{org89}Niels Bohr Institute, University of Copenhagen, Copenhagen, Denmark
\item \Idef{org90}Nikhef, National institute for subatomic physics, Amsterdam, Netherlands
\item \Idef{org91}NRC Kurchatov Institute IHEP, Protvino, Russia
\item \Idef{org92}NRNU Moscow Engineering Physics Institute, Moscow, Russia
\item \Idef{org93}Nuclear Physics Group, STFC Daresbury Laboratory, Daresbury, United Kingdom
\item \Idef{org94}Nuclear Physics Institute of the Czech Academy of Sciences, \v{R}e\v{z} u Prahy, Czech Republic
\item \Idef{org95}Oak Ridge National Laboratory, Oak Ridge, Tennessee, United States
\item \Idef{org96}Petersburg Nuclear Physics Institute, Gatchina, Russia
\item \Idef{org97}Physics department, Faculty of science, University of Zagreb, Zagreb, Croatia
\item \Idef{org98}Physics Department, Panjab University, Chandigarh, India
\item \Idef{org99}Physics Department, University of Jammu, Jammu, India
\item \Idef{org100}Physics Department, University of Rajasthan, Jaipur, India
\item \Idef{org101}Physikalisches Institut, Eberhard-Karls-Universit\"{a}t T\"{u}bingen, T\"{u}bingen, Germany
\item \Idef{org102}Physikalisches Institut, Ruprecht-Karls-Universit\"{a}t Heidelberg, Heidelberg, Germany
\item \Idef{org103}Physik Department, Technische Universit\"{a}t M\"{u}nchen, Munich, Germany
\item \Idef{org104}Research Division and ExtreMe Matter Institute EMMI, GSI Helmholtzzentrum f\"ur Schwerionenforschung GmbH, Darmstadt, Germany
\item \Idef{org105}Rudjer Bo\v{s}kovi\'{c} Institute, Zagreb, Croatia
\item \Idef{org106}Russian Federal Nuclear Center (VNIIEF), Sarov, Russia
\item \Idef{org107}Saha Institute of Nuclear Physics, Kolkata, India
\item \Idef{org108}School of Physics and Astronomy, University of Birmingham, Birmingham, United Kingdom
\item \Idef{org109}Secci\'{o}n F\'{\i}sica, Departamento de Ciencias, Pontificia Universidad Cat\'{o}lica del Per\'{u}, Lima, Peru
\item \Idef{org110}Shanghai Institute of Applied Physics, Shanghai, China
\item \Idef{org111}Stefan Meyer Institut f\"{u}r Subatomare Physik (SMI), Vienna, Austria
\item \Idef{org112}SUBATECH, IMT Atlantique, Universit\'{e} de Nantes, CNRS-IN2P3, Nantes, France
\item \Idef{org113}Suranaree University of Technology, Nakhon Ratchasima, Thailand
\item \Idef{org114}Technical University of Ko\v{s}ice, Ko\v{s}ice, Slovakia
\item \Idef{org115}Technische Universit\"{a}t M\"{u}nchen, Excellence Cluster 'Universe', Munich, Germany
\item \Idef{org116}The Henryk Niewodniczanski Institute of Nuclear Physics, Polish Academy of Sciences, Cracow, Poland
\item \Idef{org117}The University of Texas at Austin, Austin, Texas, United States
\item \Idef{org118}Universidad Aut\'{o}noma de Sinaloa, Culiac\'{a}n, Mexico
\item \Idef{org119}Universidade de S\~{a}o Paulo (USP), S\~{a}o Paulo, Brazil
\item \Idef{org120}Universidade Estadual de Campinas (UNICAMP), Campinas, Brazil
\item \Idef{org121}Universidade Federal do ABC, Santo Andre, Brazil
\item \Idef{org122}University College of Southeast Norway, Tonsberg, Norway
\item \Idef{org123}University of Cape Town, Cape Town, South Africa
\item \Idef{org124}University of Houston, Houston, Texas, United States
\item \Idef{org125}University of Jyv\"{a}skyl\"{a}, Jyv\"{a}skyl\"{a}, Finland
\item \Idef{org126}University of Liverpool, Department of Physics Oliver Lodge Laboratory , Liverpool, United Kingdom
\item \Idef{org127}University of Tennessee, Knoxville, Tennessee, United States
\item \Idef{org128}University of the Witwatersrand, Johannesburg, South Africa
\item \Idef{org129}University of Tokyo, Tokyo, Japan
\item \Idef{org130}University of Tsukuba, Tsukuba, Japan
\item \Idef{org131}Universit\'{e} Clermont Auvergne, CNRS/IN2P3, LPC, Clermont-Ferrand, France
\item \Idef{org132}Universit\'{e} de Lyon, Universit\'{e} Lyon 1, CNRS/IN2P3, IPN-Lyon, Villeurbanne, Lyon, France
\item \Idef{org133}Universit\'{e} de Strasbourg, CNRS, IPHC UMR 7178, F-67000 Strasbourg, France, Strasbourg, France
\item \Idef{org134} Universit\'{e} Paris-Saclay Centre d¿\'Etudes de Saclay (CEA), IRFU, Department de Physique Nucl\'{e}aire (DPhN), Saclay, France
\item \Idef{org135}Universit\`{a} degli Studi di Pavia, Pavia, Italy
\item \Idef{org136}Universit\`{a} di Brescia, Brescia, Italy
\item \Idef{org137}V.~Fock Institute for Physics, St. Petersburg State University, St. Petersburg, Russia
\item \Idef{org138}Variable Energy Cyclotron Centre, Kolkata, India
\item \Idef{org139}Warsaw University of Technology, Warsaw, Poland
\item \Idef{org140}Wayne State University, Detroit, Michigan, United States
\item \Idef{org141}Westf\"{a}lische Wilhelms-Universit\"{a}t M\"{u}nster, Institut f\"{u}r Kernphysik, M\"{u}nster, Germany
\item \Idef{org142}Wigner Research Centre for Physics, Hungarian Academy of Sciences, Budapest, Hungary
\item \Idef{org143}Yale University, New Haven, Connecticut, United States
\item \Idef{org144}Yonsei University, Seoul, Republic of Korea
\end{Authlist}
\endgroup
\end{document}